\documentclass[fleqn,usenatbib]{mnras}

\usepackage[T1]{fontenc}

\DeclareRobustCommand{\VAN}[3]{#2}
\let\VANthebibliography\thebibliography
\def\thebibliography{\DeclareRobustCommand{\VAN}[3]{##3}\VANthebibliography}

\usepackage{graphicx}	
\usepackage{amsmath}	
\usepackage{amssymb}	
\usepackage{subfigure}

\usepackage{newtxtext,newtxmath}

\title[Spiral Density Enhancements in Be Binary Systems]{Spiral Density Enhancements in Be Binary Systems}

\author[I. H. Cyr et al.]{
Isabelle. H. Cyr,$^{1}$
C. E. Jones,$^{1}$
A. C. Carciofi,$^{2}$
C. Steckel,$^{1}$
C. Tycner,$^{3}$
A. T. Okazaki$^{4}$
\\
$^{1}$Department of Physics and Astronomy, The University of Western Ontario, London, ON Canada N6A 3K7\\
$^{2}$Instituto de Astronomia, Geof\'isica e Ci\^encias Atmosf\'ericas, Universidade de S\~ao Paulo, S\~ao Paulo, Brazil\\
$^{3}$Department of Physics, Central Michigan University, Mount Pleasant, MI, USA 48859\\
$^{4}$Hokkai-Gakuen University, Toyohira-ku, 062-8605, Sapporo, Japan
}

\date{Accepted XXX. Received YYY; in original form ZZZ}

\pubyear{2020}

\begin{document}
\label{firstpage}
\pagerange{\pageref{firstpage}--\pageref{lastpage}}
\maketitle

\begin{abstract}
We use a smoothed particle hydrodynamics (SPH) code to examine the effects of a binary companion on a Be star disk for a range of disk viscosities and misalignment angles, i.e. the angle between the orbital plane and the primary's spin axis. The density structures in the disk due to the tidal interaction with the binary companion are investigated. Expanding on our previous work, the shape and density structure of density enhancements due to the binary companion are analyzed and the changes in observed interferometric features due to these orbiting enhancements are also predicted. We find that larger misalignment angles and viscosity values result in more tightly wound spiral arms with densities that fall-off more slowly with radial distance from the central star. We show that the orbital phase has very little effect on the structure of the spiral density enhancements. We demonstrate that these spiral features can be detected with an interferometer in H$\alpha$ and K-band emission. We also show that the spiral features affect the axis ratios determined by interferometry depending on the orientation of these features and the observer. For example, our simulations show that the axis ratios can vary by 20\% for our co-planar binary disk system depending on the location of the disk density enhancements.
\end{abstract}




\section{Introduction}
\label{intro}

The formal definition of a Be star, developed in its current form by \citet{col87}, is a ``non-supergiant B star whose spectrum has, or had at some time, one or more Balmer lines in emission." This emission originates in a geometrically thin circumstellar disk formed from gas ejected from the rapidly rotating central star. Other important properties of these stars include, for example, infrared and radio excess, and intrinsic linear polarization due to radiative processes within the disk.  

The formal definition implies that variability is an inherent property of these stars. The observed variability occurs on a variety of time scales from short-term spectral variations on periods of minutes to hours, thought to be due to stellar pulsations or changes in the inner disk \citep{baa16}, to periods of time scales of an order of decades associated with the complete loss or renewal of the disk \citep{wis10}. 

Intermediate periods are associated with the variations in the ratio of the violet to red peaks (V/R ratios) of doubly peaked emission lines. It has long been suggested that the origin of the V/R variations is due to rotating density enhancements within the disk. Large scale, one-arm orbiting density waves operate on periods of years \citep{ste09} and shorter term, smaller amplitude variations associated with binary companions with periods of weeks to months have been reported. Given that V/R variations occur in approximately 2/3 of Be stars \citep[see][]{kog82}, these features represent an important characteristic that needs to be fully understood. Realistic dynamic models that focus on the time evolution of the disk structure will be required to be able to account for V/R variability. However, the detailed structure of these rotating enhancements has not been thoroughly investigated and this is the motivation for this study.

Historically, \citet{str31} proposed an elegant, axisymmetric disk model that explained the formation of the disk by ejected stellar material due to rapid rotation of the central star, and he attributed the variety of the different line profile shapes to the angle between the plane of the disk and the observer. However, observations of V/R variations revealed that this model was too simplistic, and \citet{hir84} mentioned that researchers at that time attributed these type of variations to stellar pulsation or precession of an elongated envelope. \citet{kat83} was one of the first to suggest that the development of a one-armed spiral structure within the Be~star disk as the source of the V/R variation. One can imagine that if the portion of the disk where there is a density enhancement is moving toward the observer's line of sight then an increase in the violet side of the line may be observed and vice versa. \citet{oka91} studied long-term V/R variations due to global one-armed oscillations within Be star disks and concluded that detailed line profile variations analyses are required to constrain models in order to understand the V/R variations \citep[see, for example, section 5.3.2 in the review paper by][where a literature summary of the observed characteristics of V/R variations are presented]{riv13}.

    In our case, however, the analysis is more complicated. \citet{san12} reports that 75$\%$ of massive stars with masses M $>$ 8~M$_{\sun}$ are binaries or were part of a binary system at some time. This means that many, if not all, Be stars could be members of a binary system. In binary systems, tidal interactions, radiative interactions, and in the case of Be/X-ray systems high energy interactions, must all be taken into account. Not only do tidal interactions truncate the disk, but a build-up of density at radial distances less than the truncation radius, called the accumulation effect, also seems to occur \citep{pan16,cyr17}.   

\citet{oka02} studied the effects on coplanar decretion disks and neutron stars in Be/X-ray binaries and found that in binary systems a pattern of double spiral arms is set up, which may indeed become phase-locked with the companion. It is not completely clear how having two rotating arms would affect the shape and variation of doubly peaked emission lines. For example, if one arm of the spiral was bigger and/or denser, it may contribute more significantly to the line profile. Although the Be/X-ray systems have been increasingly well studied, the literature reveals much less detailed studies on normal Be star binary systems.

The periastron passage in an eccentric binary Be star system, $\delta$ Sco, provided an opportunity to study the effects on the disk by the secondary companion during the closest approach~\citep{tyc11}. \citet{che12} studied this system based on long-baseline interferometric observations that clearly resolved the binary components even at the closest approach, and their work revealed apparent asymmetries in the disk around the primary companion, but due to limited angular resolution they were not able to conclusively determine any well-defined patterns. The same system was also monitored with polarization observations by \citet{bed12}, and these authors reported that there were significant changes occurring in the surface density of the disk during the same periastron passage. 

\citet{pan16} studied Be binary systems in co-planar orbits to determine how the disk structure is altered in binary systems. They found the most significant effects are the truncation of the disk and the accumulation of disk material inside the truncation radius. In a follow-up paper, \citet{cyr17}, hereafter Paper 1, studied the effects on the disk due to the presence of an orbiting companion at a variety of disk viscosities, misalignment angles, i.e. the angle between the orbital plane and the primary's spin axis and orbital periods. They show that these parameters most significantly affect the truncation radius and the outer portion of the disk, while the innermost disk remains unaffected. More recently, \citet{pan17} investigated phased-locked changes in emission lines due to tidally locked rotating spiral-shaped density enhancements. They demonstrate that these density enhancements result in phase-locked V/R ratios and that this could be an effective tool to investigate binary disk systems.  Interestingly, they show that triply-peaked emission lines can be formed in co-planar systems without disk warping. Later, this work was followed up by \citet{pan19} who investigated disk variability due to spiral density enhancements on polarization signature and photometry for binary systems, and they found that there was a relationship between the V/R variations and polarization. Perhaps this not surprising since the rotation of the enhanced densities in the spiral arms will affect both of these observed features. Most recently, \citet{bro19} studied the disk density and size for Be/X-ray binaries. They find that not only the disk density is related to the disk viscosity and mass ejection rate, but that the disk size is strongly related to the orbital period of the companion. 

In this work, as a follow-up to Paper~1, we investigate the size and shape of disk density structures for a Be binary system. We also examine how these features change with viscosity, orbital phase, and misalignment angle. Finally, we predict how spiral density enhancements can affect interferometric measurements of overall disk dimensions. This work is organized as follows: details about our simulations are provided in Section~\ref{method}, our results are presented in Section~\ref{sec4:results}, and a summary and discussion can be found in Section~\ref{sec4:conclusion}.

\section{Methodology}
\label{method}

In this work, we use the SPH simulations described in Paper 1 to probe the structure of the spiral arms that develop in the disk of Be binary systems. Our models are based on  viscous decretion disk, VDD, theory originally adopted by \citet{lee91} for Be star disks following the prescription for accretion disks developed by \citet{sha73}. Our SPH simulations approach the hydrodynamic solution with the viscosity parameterized by the standard $\alpha$ viscosity \citep{sha73}, denoted by $\alpha_\mathrm{SS}$ in this study. Basically, we start with a disk-less system and inject equal mass particles in Keplerian orbit. The injected particles interact and transfer angular momentum outward although approximately 98\% of the injected particles fall back onto the star. We note that whenever particles enter an accretion radius of the primary or secondary, they are removed from the simulation. Our simulated disks are isothermal and we adopt a disk temperature 60\% of the stellar effective temperature following the suggestions of \citet{mil98} and \citet{car06}. Our analysis of the spiral density enhancements follows a total of 50 orbital periods.

Following a similar approach to our earlier work, we investigate the effects of the following three parameters on the structure of the arms; the viscosity of the disk ($\alpha_\mathrm{SS}$ = 0.1, 0.5, 1.0) the misalignment angle ($\theta$ = 0$\degr$, 30$\degr$, 45$\degr$, 60$\degr$), and the orbital phase ($p$ = 0.25, 0.50, 0.75, 1.00). We note that for $p$ = 0.50 and 1.00 (or zero) the binary companion crosses the equatorial plane and for $p$ = 0.25 and 0.50 the companion is below and above the equatorial plane, respectively. See figure~1 in \citet{cyr17} for a pictorial representation of these details. The surface density of the disk,  $\Sigma(r,\phi)$, as a function of radial distance, $r$, and azimuthal angle, $\phi$, for this binary system is shown in Figure~\ref{spiral and phases} along with the orientation of the phases, $p$ of 0.25, 0.50, 0.75, 1.00, in panels labeled a) to d), respectively. As in Paper 1, we define $p$ = 0.00 and 1.00 to be the the beginning of 49$^{th}$ and 50$^{th}$ orbital period ($P_{orb}$), respectively, where the disk has reached steady state. The stellar parameters were based loosely on the Pleione binary system \citep[see][]{har88,hir07,nem10}. However, the main goal of this study is not specifically focused on this particular binary system, but rather on exploring the the nature of the spiral structures and their potential impact on interferometric observables. Therefore, several parameters were modified in order to simplify the system for use in our investigations. For convenience we provide the adopted stellar and orbital parameters for our simulated system in Table~\ref{parameters}. Note that we restricted our study to simulations with an orbital period of 30 days, which corresponds to a semi-major axis of 59.5 $R_\odot$. For additional details about the SPH code and these simulations, see \citet{oka02} and Paper 1.

\begin{table}
\centering
\caption{Adopted parameters of the binary system.}
\label{parameters}
\begin{tabular}{lc}
\hline
Parameter&Value\\
\hline
Mass of primary & 2.9 M$_{\sun}$\\ 
Radius of primary & 3.67 R$_{\sun}$\\ 
Effective temperature & 12,000 K\\
 & \\
Mass of secondary & 0.31 M$_{\sun}$\\
Radius of secondary & 0.38 R$_{\sun}$\\
 & \\
Mass injection rate & 10$^{-8}$ M$_{\sun}$ yr$^{-1}$\\
\hline
\end{tabular}
\end{table}

Output files from our SPH simulations were adopted for the 3D radiative transfer code, \textsc{hdust}, originally developed and presented in \citet{car06}. This code has been used extensively to predict observables, including emission line profiles, spectral energy distributions, interferometry and polarization from Be disk systems \cite[see the review by][and references therein]{riv13}.

Specifically, we formatted the SPH output files so that \textsc{hdust} could be used to produce images to show the specific intensity in wavelength bands of interest. Once we constructed an image in a particular wavelength region, we computed a 2D Fourier transform of the image. The Fourier transforms (FT) were computed using the fftpack module of Python's SciPy library, which uses the Cooley-Tukey fast Fourier transform algorithm \citep{vir19}. 

To simulate what would typically be observed by a long-baseline interferometer with a single baseline with a fixed orientation with respect to east-west and north-south directions, we took cross-sections of the Fourier transformed image (computed as a normalized Fourier power) at different angular orientations (in one degree steps) as measured from the spatial frequency axis that corresponds to the x-axis in our image plane.  The resulting normalized Fourier power (also known as the squared visibility) as a function of radial spatial frequency was subsequently fit with Gaussian function that allowed us to obtain a single full-width at half-maximum~(FWHM) measure of the disk structure along a specific orientation with respect to the x-axis.  By dividing two Gaussian FWHM measures from two orthogonal directions, allowed us to compute an effective axis ratio that would be produced based on a two-baseline interferometer (with orthogonal baselines projected on the sky).

This procedure was validated prior to our analysis of our disk models with spiral density enhancements by applying the same analysis to fully circular disks, as well as elliptical disks with specific (and known) axis ratios. The FT of a uniform circular flux distribution produced the expected axis ratios that were unity for all orthogonal FT cross-cuts, regardless of the orientation angle. The FT of an elliptical brightness distribution produced narrower or wider FWHM Gaussian measures, depending on the orientation of the cross-sections in the FT space.  By exploring orientations across two quadrants in the Fourier space (i.e., across 180$^{\circ}$) we were able to recover the appropriate $a/b$ axis ratio that is expected for elliptical brightness distributions.

\section{Results}
\label{sec4:results}
\subsection{Shape of Density Enhancements}
\begin{figure*}
\center
\subfigure[$p = 0.25$]{\includegraphics[width=0.45\textwidth]{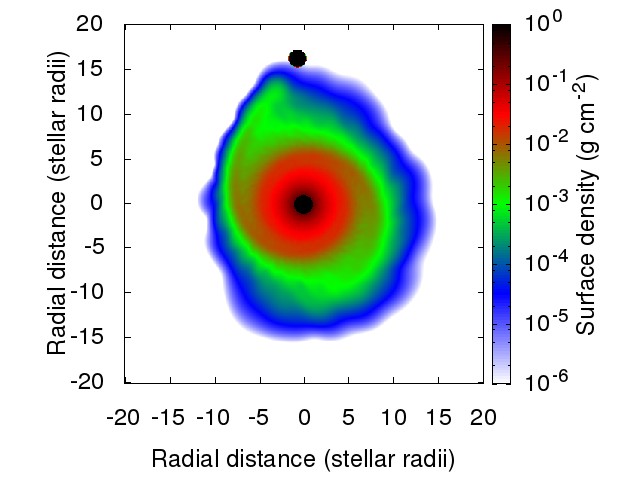}} 
\subfigure[$p = 0.50$]{\includegraphics[width=0.45\textwidth]{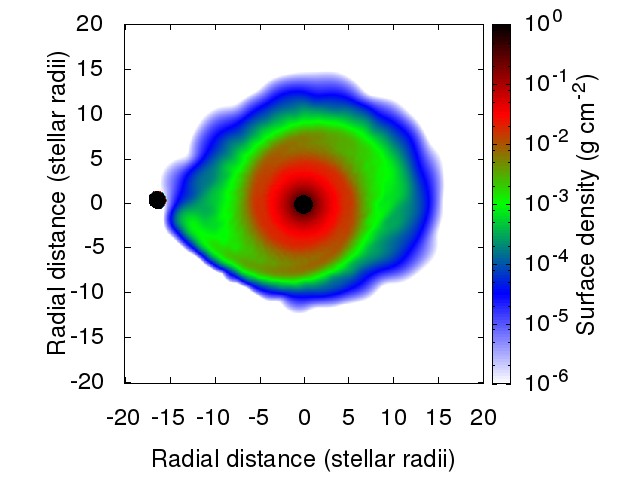}} \\
\subfigure[$p = 0.75$]{\includegraphics[width=0.45\textwidth]{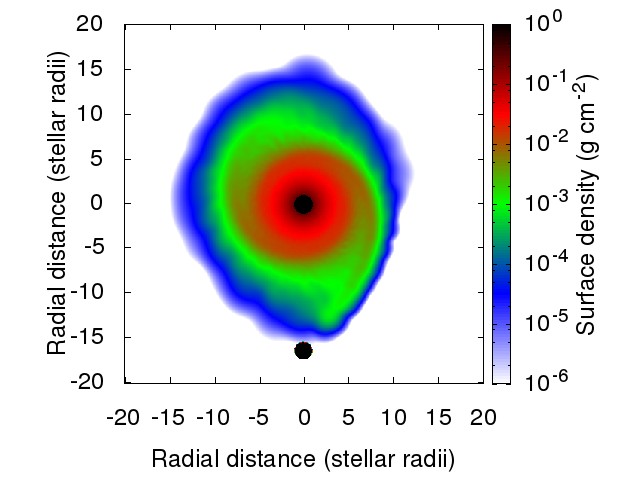}} 
\subfigure[$p = 1.00$]{\includegraphics[width=0.45\textwidth]{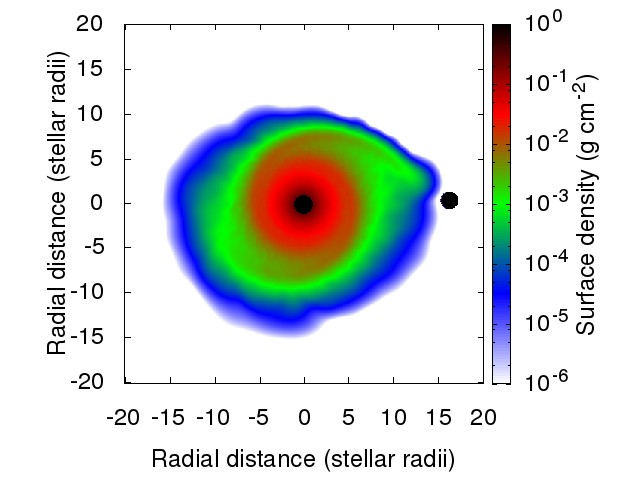}} \\
\caption[]{Map of the surface density, $\Sigma(r,\phi)$, of an aligned disk ($\theta$ = 0$\degr$) at orbital phases of (a) $p$ = 0.25, (b) $p$ = 0.50, (c) $p$ = 0.75, and (d) $p$ = 1.00. The black circles are the central star and binary companion. The black circle at the edge of disk is the secondary star that orbits the primary and disk counter-clockwise in this top-down view. The arm ahead of the secondary is referred to as the leading arm and the arm following the secondary's orbit we call the trailing arm. Viscosity is set to $\alpha_\mathrm{SS}$ = 0.5 for this Figure.}
\label{spiral and phases}
\end{figure*}

Figure~\ref{spiral and phases} shows a top-down view of $\Sigma(r,\phi)$ for one of our representative simulations with $\alpha_\mathrm{SS} = 0.5$ and $\theta = 0\degr$ seen for $p$ of 0.25, 0.5, 0.75, and 1.0. The primary and secondary stars are represented by the black circles. See figure~1 in Paper 1 for a diagram showing more details of the coordinate system. 

We can clearly see the presence of two density enhancements, or spiral arms, winding around the central star. We note that the secondary star is orbiting the primary in a counter-clockwise direction when viewing Figure~\ref{spiral and phases} top-down. The arm that is ahead of the orbit of the secondary, we call the leading arm and we refer to the arm following the secondary, the trailing arm. By comparing the position of these arms at different phases we see that they appear to be tidally locked with the secondary, with one arm always leading toward the same side as the secondary (leading arm) and the other trailing on the opposite side (trailing arm). This holds true for all values of viscosity and misalignment angles that we considered.

In order to describe the shape of these density enhancements, we investigated the surface density as a function of the azimuthal angle, $\phi$, at different fixed radial distances, $r$, which we will refer to as $\Sigma_r(\phi)$. Then we used a series of Gaussian functions to fit the surface density profiles to determine the central position of the arm along the azimuthal axis ($\phi_r^0$) for each $r$,

\begin{equation}
\Sigma_r(\phi) = \left (\Delta\Sigma_r \right) e^{-(\phi - \phi_r^0)^2 / \sigma_r^2 } + \Sigma_r^0,
\label{eq4:sigma_phi}
\end{equation}
where $\Sigma_r^0$ is the ambient surface density of the disk as a function of radial distance, $r$, i.e. the surface density outside of the location of density enhancements, $\Delta\Sigma_r$ is the maximum surface density enhancement, and $\sigma_r$ is the width of the enhancement. The subscript \textit{r} represents the radial distance so that Equation~\ref{eq4:sigma_phi} is applied systematically in concentric rings as \textit{r} increases. The surface density of the arms is written as,
\begin{equation}
\Sigma_{arm} = \Sigma_r^0 + \Delta\Sigma_r.
\label{eq4:sigma_arm}
\end{equation}

\begin{figure*}
\center
\subfigure[Leading arm]{\includegraphics[width=0.45\textwidth,keepaspectratio]{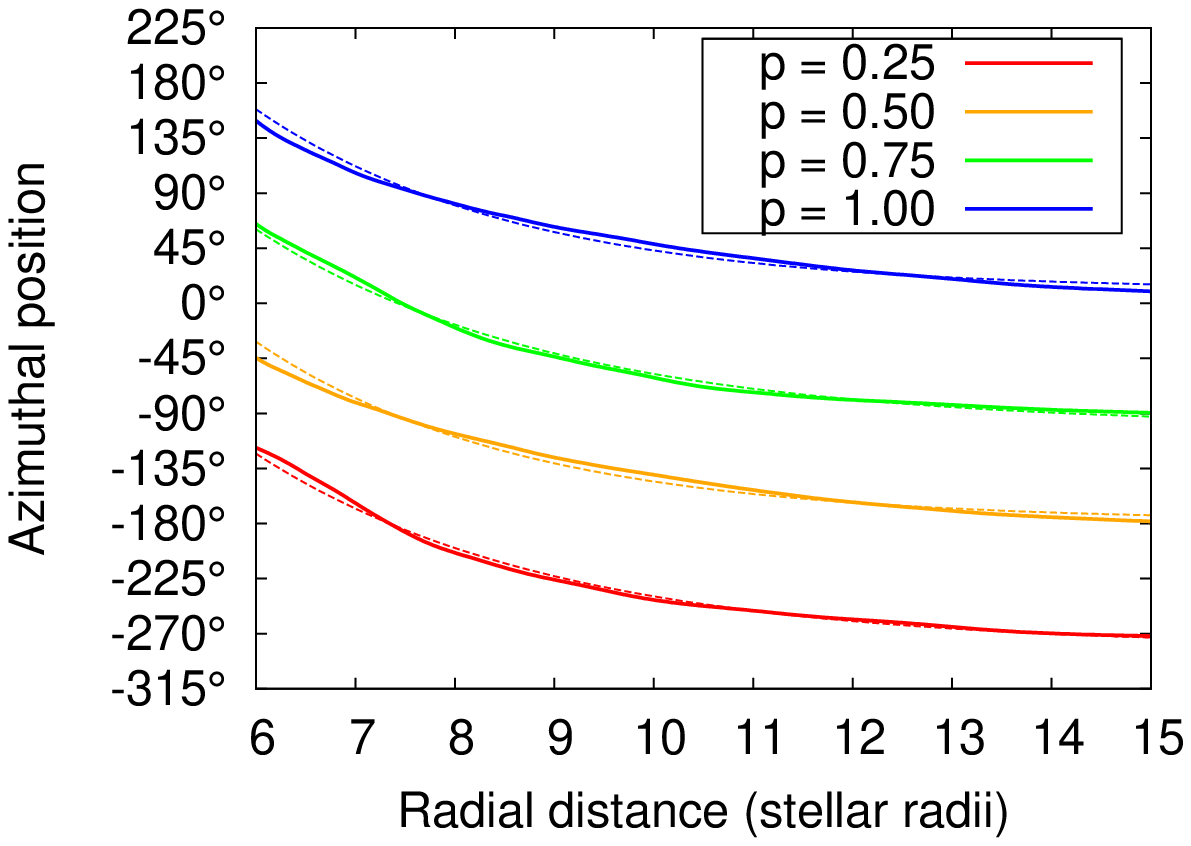}} 
\subfigure[Trailing arm]{\includegraphics[width=0.45\textwidth,keepaspectratio]{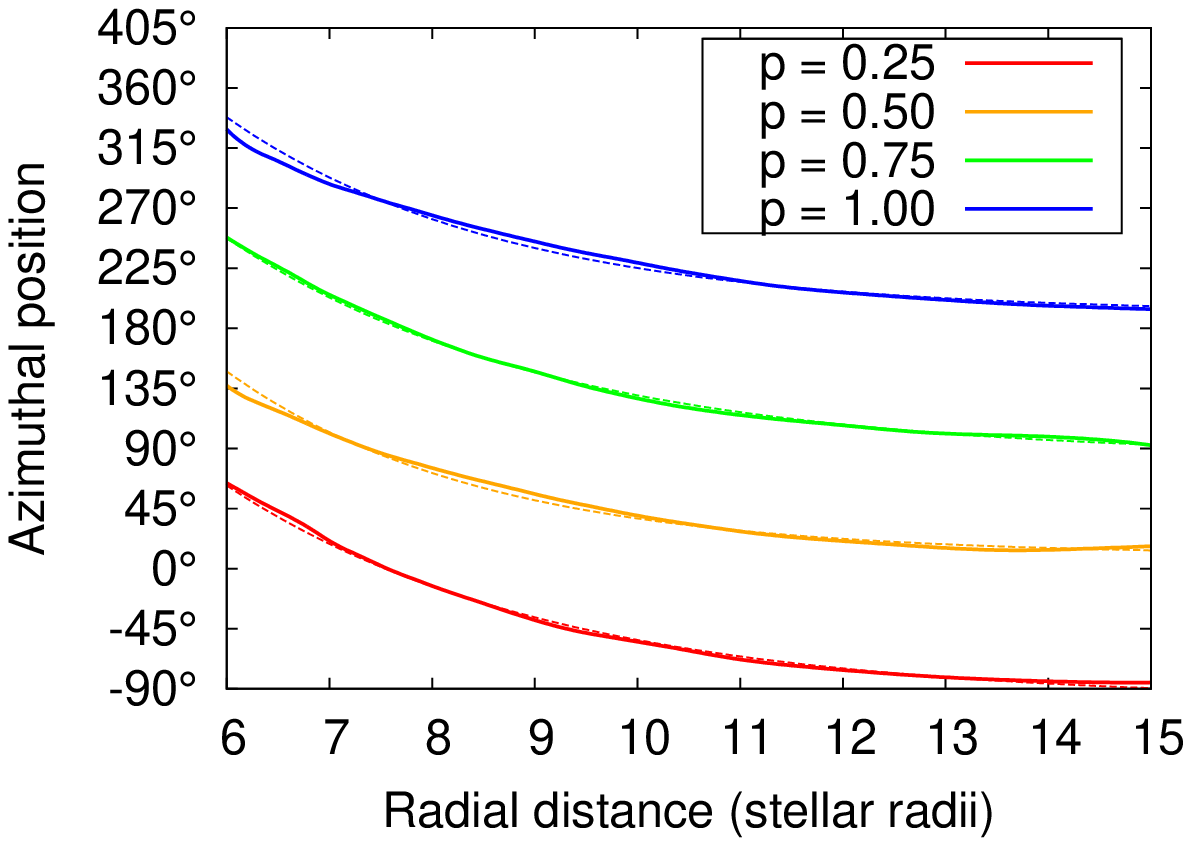}} 
\caption[Azimuthal position of spiral arms as a function of $r$]{Azimuthal position of the leading (a) and trailing (b) spiral arm as a function of radial distance, at orbital phases of $p$ = 0.25 (red), 0.50 (orange), 0.75 (green), and 1.00 (blue). Results obtained from our simulations are shown as solid lines while the results of the exponential fit, described by Equation~\ref{eq4:phi_r}, are represented by the dashed lines.}
\label{fig4:phi_r}
\end{figure*}

Figure~\ref{fig4:phi_r} shows the azimuthal position, $\phi_0$, of the leading and trailing arms as a function of radial distance obtained from Equation~\ref{eq4:sigma_phi}. We notice that as the phase of the orbit increases, the azimuthal position of each arm shifts by 90$\degr$ as expected since the motion of the arms is locked with the secondary. We also notice a shift of approximately 180$\degr$ between leading and trailing arms, which is also expected since the arms are situated approximately on opposite sides of the primary.

In order to better describe the shape of these spiral features, we fit each density profile using an exponential function,
\begin{equation}
\phi(r) = A e^{-\gamma r} + B,
\label{eq4:phi_r}
\end{equation}
where $A$ and $B$ are fitting constants, and $\gamma$ is a parameter related to the winding of the spiral arms. Smaller values of $\gamma$ indicate tighter winding. The values of $\gamma$ over the full range of model parameters are provided in Tables~\ref{table4:fit_leading} and \ref{table4:fit_trailing}, for the leading and trailing arms, respectively.

\begin{table}
\centering
\caption{Winding parameter $\gamma$ of the leading arms}
\begin{tabular}{cccccc}
\hline
\hline
Viscosity & Misalignment & \multicolumn{4}{c}{{{Orbital phase [$p$]}}}\\
\cline{3-6}
\noalign{\vskip 0.1cm} 
parameter [$\alpha_\mathrm{SS}$] & angle [$\theta$] & 0.25 & 0.50 & 0.75 & 1.0\\
\hline

 & 0$\degr$ & 0.493 & 0.494 & 0.487 & 0.489 \\
0.1 & 30$\degr$ & 0.426 & 0.469 & 0.450 & 0.465 \\
 & 45$\degr$ & 0.377 & 0.448 & 0.370 & 0.443 \\
 & 60$\degr$ & 0.336 & 0.415 & 0.357 & 0.413 \\
\hline
\noalign{\vskip 0.1cm} 
 & 0$\degr$ & 0.400 & 0.385 & 0.388 & 0.378 \\
0.5 & 30$\degr$ & 0.334 & 0.377 & 0.330 & 0.378 \\
 & 45$\degr$ & 0.294 & 0.370 & 0.278 & 0.362 \\
 & 60$\degr$ & 0.281 & 0.367 & 0.281 & 0.357 \\
\hline
\noalign{\vskip 0.1cm} 
 & 0$\degr$ & 0.340 & 0.341 & 0.368 & 0.393 \\
1.0 & 30$\degr$ & 0.319 & 0.360 & 0.270 & 0.328 \\
 & 45$\degr$ & 0.270 & 0.362 & 0.269 & 0.274 \\
 & 60$\degr$ & 0.249 & 0.318 & 0.239 & 0.332 \\

\hline
\end{tabular}
\label{table4:fit_leading}
\end{table}

\begin{table}
\centering
\caption{Winding parameter $\gamma$ of the trailing arms}
\begin{tabular}{cccccc}
\hline
\hline
Viscosity & Misalignment & \multicolumn{4}{c}{{{Orbital phase [$p$]}}}\\
\cline{3-6}
\noalign{\vskip 0.1cm} 
parameter [$\alpha_\mathrm{SS}$] & angle [$\theta$] & 0.25 & 0.50 & 0.75 & 1.0\\
\hline

 & 0$\degr$ & 0.503 & 0.456 & 0.439 & 0.449 \\
0.1 & 30$\degr$ & 0.389 & 0.505 & 0.403 & 0.485 \\
 & 45$\degr$ & 0.391 & 0.406 & 0.389 & 0.433 \\
 & 60$\degr$ & 0.342 & 0.388 & 0.334 & 0.383 \\
\hline
\noalign{\vskip 0.1cm} 
 & 0$\degr$ & 0.391 & 0.366 & 0.355 & 0.387 \\
0.5 & 30$\degr$ & 0.316 & 0.402 & 0.315 & 0.366 \\
 & 45$\degr$ & 0.283 & 0.358 & 0.269 & 0.384 \\
 & 60$\degr$ & 0.285 & 0.300 & 0.278 & 0.296 \\
\hline
\noalign{\vskip 0.1cm} 
 & 0$\degr$ & 0.442 & 0.342 & 0.358 & 0.350 \\
1.0 & 30$\degr$ & 0.308 & 0.294 & 0.280 & 0.326 \\
 & 45$\degr$ & 0.253 & 0.315 & 0.272 & 0.323 \\
 & 60$\degr$ & 0.281 & 0.226 & 0.281 & 0.275 \\

\hline
\end{tabular}
\label{table4:fit_trailing}
\end{table}

\begin{figure*}
\center
\subfigure[]{\includegraphics[width=0.45\textwidth]{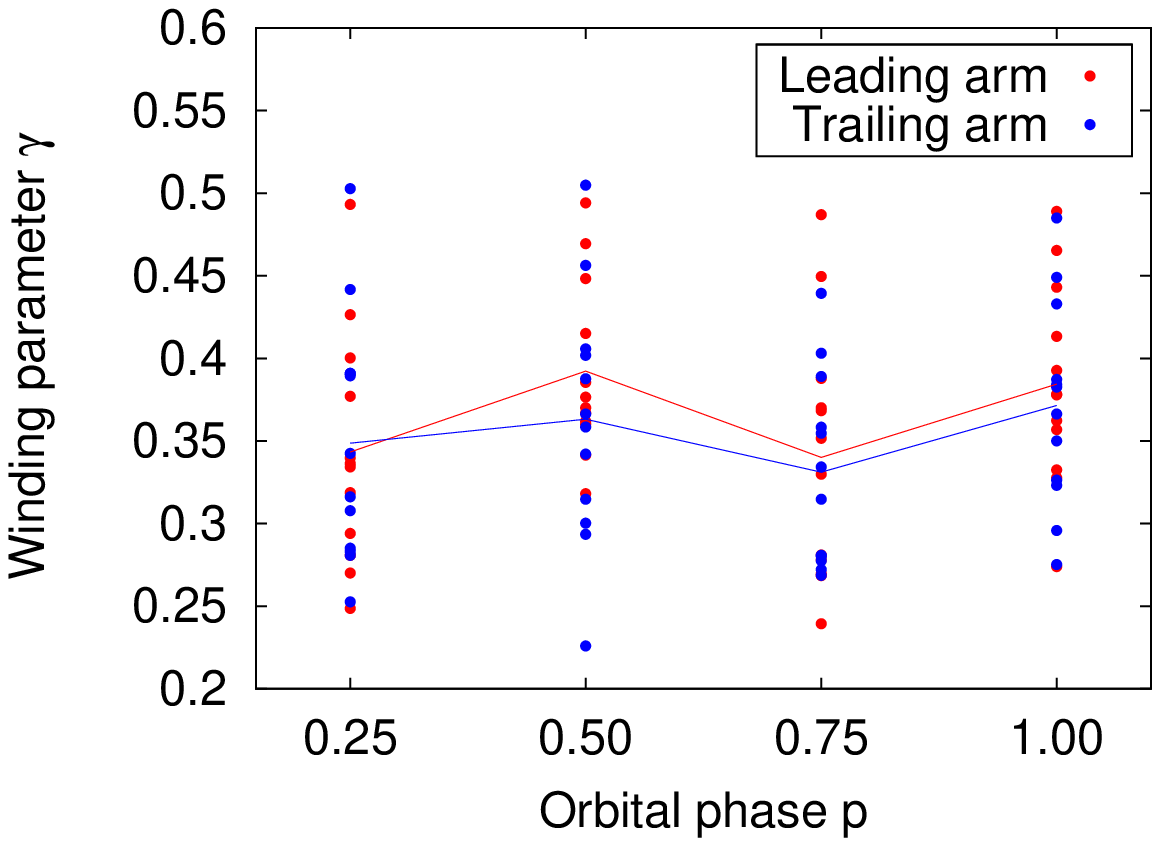}} 
\subfigure[]{\includegraphics[width=0.45\textwidth]{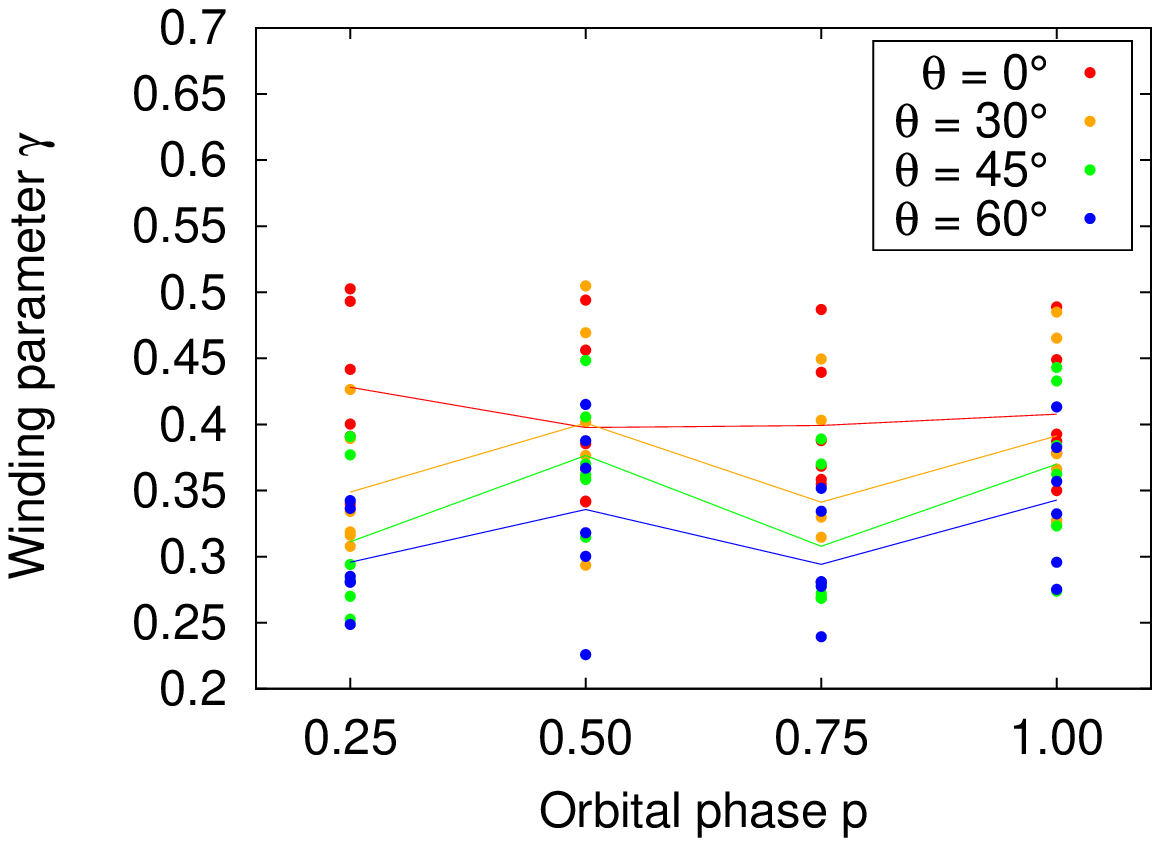}}
\caption[Winding parameter of the spiral arms as a function of orbital phase]{Winding parameter, $\gamma$, of spiral arms as a function of orbital phase, $p$. In panel (a), values are grouped into leading arms (red) and trailing arms (blue) while in panel (b) the values are grouped by misalignment angle; 0$\degr$ (red), 30$\degr$ (orange), 45$\degr$ (green), and 60$\degr$ (blue). The average of each of these groups as a function of $p$ is represented by the solid lines.}
\label{fig4:b_ph}
\end{figure*}

Figure~\ref{fig4:b_ph} shows the values of $\gamma$ for all 96 models as a function of orbital phase. The spread of $\gamma$ values is nearly the same for all four orbital phases, suggesting that the orbital phase plays a minor role in the winding of the disk. However, small variations can be seen in the average $\gamma$ values (shown with solid lines in Figure~\ref{fig4:b_ph}). 
From Figure~\ref{fig4:b_ph}(a), we see that the winding of both arms are very similar, with trailing arms being slightly more tightly wound compared to the leading arms. Interestingly, Figure~\ref{fig4:b_ph}(b) shows that these small variations seen in panel (a) originate mainly from misaligned systems ($\theta$ = 30$\degr$, 45$\degr$, and 60$\degr$), while the aligned systems show almost no dependence on $p$. Furthermore, these variations seem to oscillate, having smaller $\gamma$ values (more tightly wound arms) at $p$ = 0.25 and 0.75 than at $p$ = 0.50 and 1.00. These features all point toward the fact that the degree of elevation of the secondary above or below to the equatorial plane has an impact on the winding of the spiral features in the disk.

\begin{figure}
\center
\includegraphics[width=0.45\textwidth]{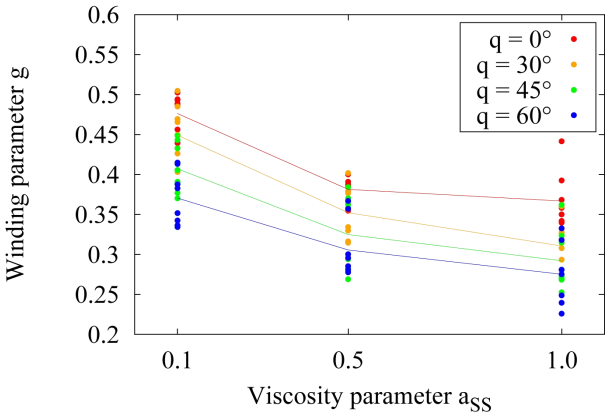}
\caption[Winding parameter of spiral arms as a function of disk viscosity]{Same as Figure~\ref{fig4:b_ph}(b) except plotted as a function of disk viscosity $\alpha_\mathrm{SS}$.}
\label{fig4:b_al}
\end{figure}

\begin{figure}
\center
\includegraphics[width=0.45\textwidth]{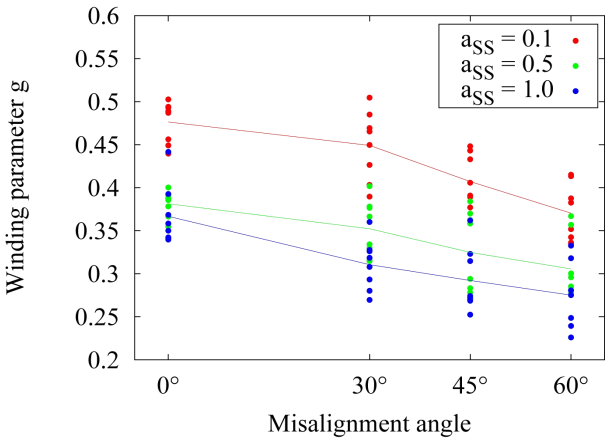}
\caption[Winding parameter of spiral arms as a function of misalignment angle]{Same as Figures~\ref{fig4:b_ph} and \ref{fig4:b_al} except values are plotted as a function of the misalignment angle, $\theta$. The colours represent different disk viscosity values; $\alpha_\mathrm{SS}$ = 0.1 (red), 0.5 (green) and 1.0 (blue). }
\label{fig4:b_th}
\end{figure}

Figure~\ref{fig4:b_al} shows the same $\gamma$ values plotted as a function of the viscosity parameter, $\alpha_\mathrm{SS}$. For this Figure, the data points were divided based on their misalignment angle in the same way as in Figure~\ref{fig4:b_ph}(b). Again, the solid lines show the average values of each set of misalignment angles. Figure~\ref{fig4:b_th} is similar to Figures~\ref{fig4:b_ph} and \ref{fig4:b_al} but with $\gamma$ plotted as a function of misalignment angle with different colours representing different $\alpha_\mathrm{SS}$ values.

Figure~\ref{fig4:b_al} demonstrates that the winding of the arms has a clear dependence on the viscosity of the disk, with smaller values of $\gamma$ for larger $\alpha_\mathrm{SS}$, i.e. the more viscous the disk, the tighter the spiral arms. This is because the greater the value of $\alpha_\mathrm{SS}$, i.e., higher viscosity, the less perturbed the disk is due to the companion, resulting in less elliptical orbits and therefore more tightly wound arms. This holds true for all four misalignment angles. Finally, we notice that the winding of the arms is also dependent on the misalignment angle, with larger $\theta$ resulting in more tightly wound arms (smaller $\gamma$). With increasing misalignment angle, the companion is less effective perturbing the disk and the result is spiral arms that are more tightly wound. Both of these dependencies can also be seen in Figure~\ref{fig4:b_th}. Interestingly, we see that the dependence on $\alpha_\mathrm{SS}$ does not appear to be linear. In contrast, we see a much more linear dependence with $\theta$.

\begin{figure*}
\center
\subfigure[$\theta = 0\degr$]
{\includegraphics[width=0.28\textheight,keepaspectratio]{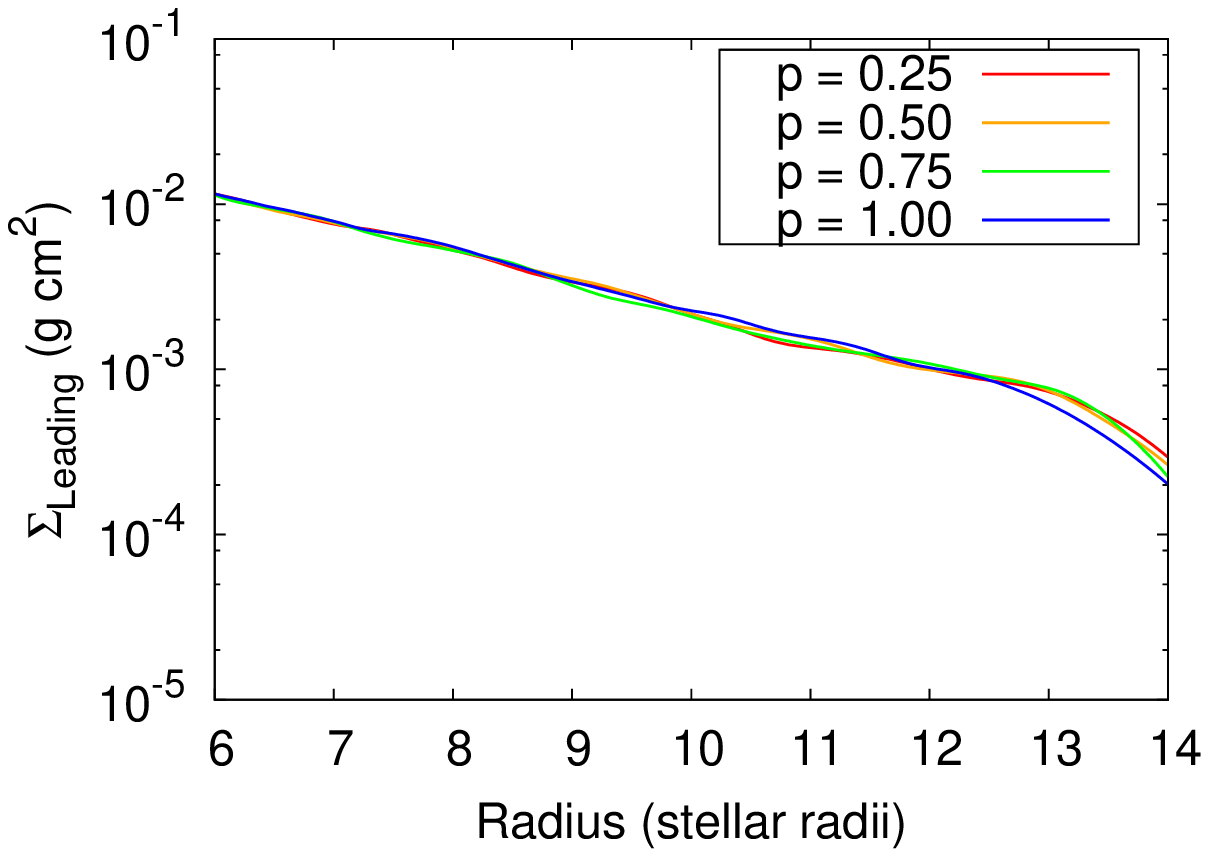}} 
\subfigure[$\theta = 0\degr$]
{\includegraphics[width=0.28\textheight,keepaspectratio]{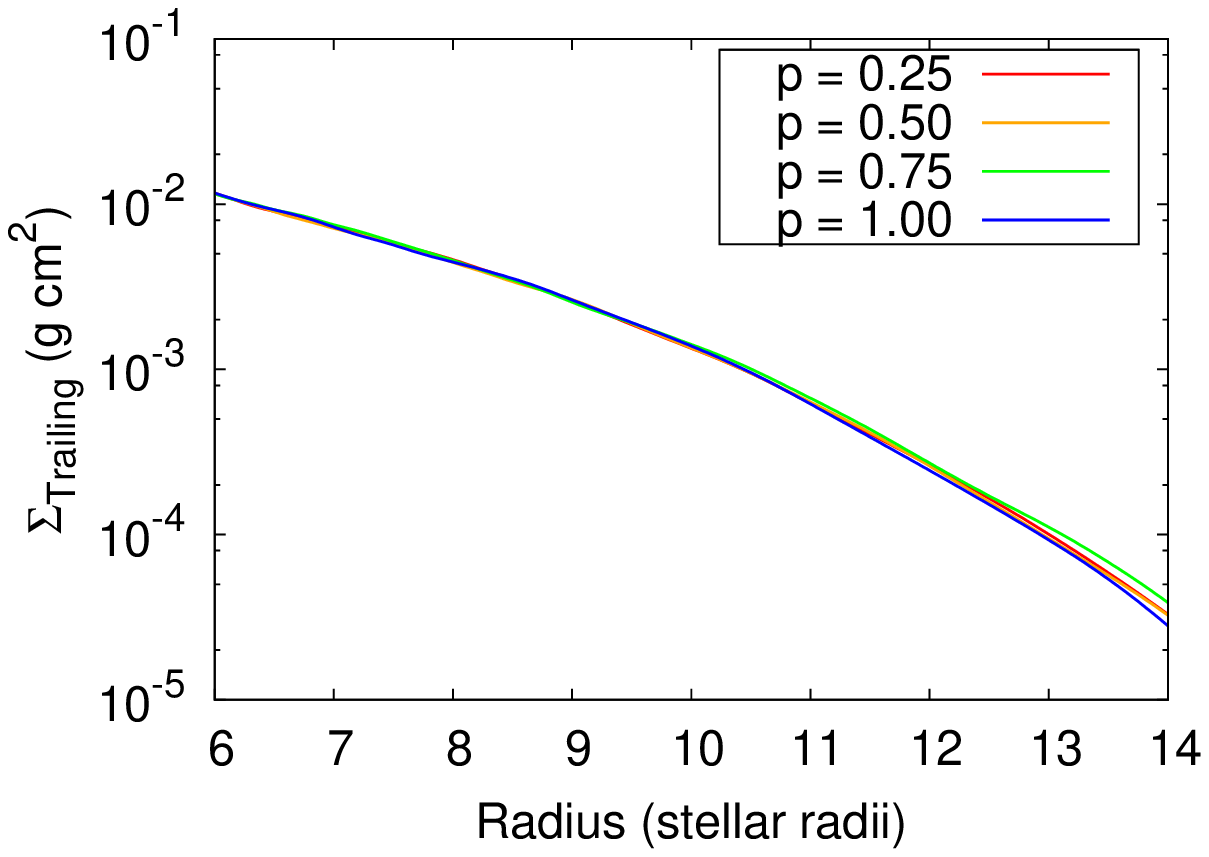}} 
\\
\subfigure[$\theta = 30\degr$]
{\includegraphics[width=0.28\textheight,keepaspectratio]{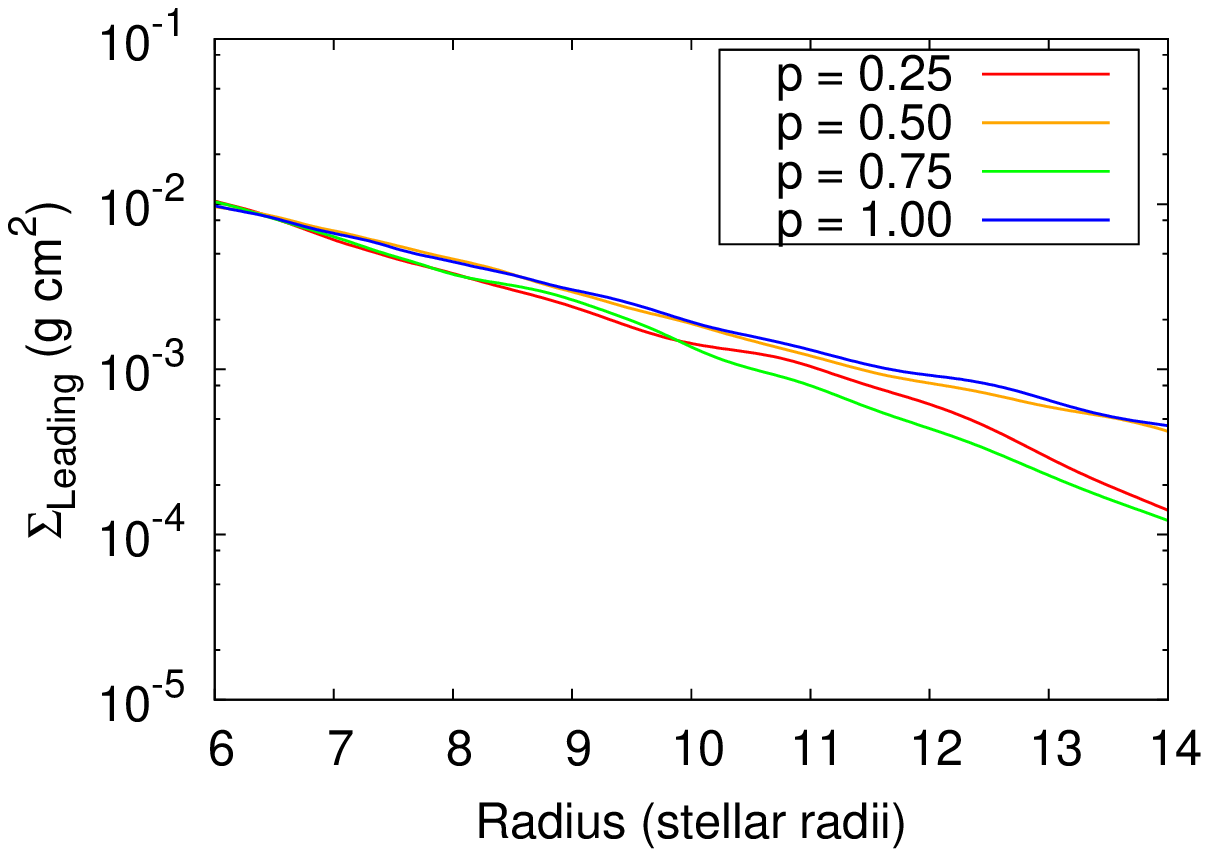}} 
\subfigure[$\theta = 30\degr$]
{\includegraphics[width=0.28\textheight,keepaspectratio]{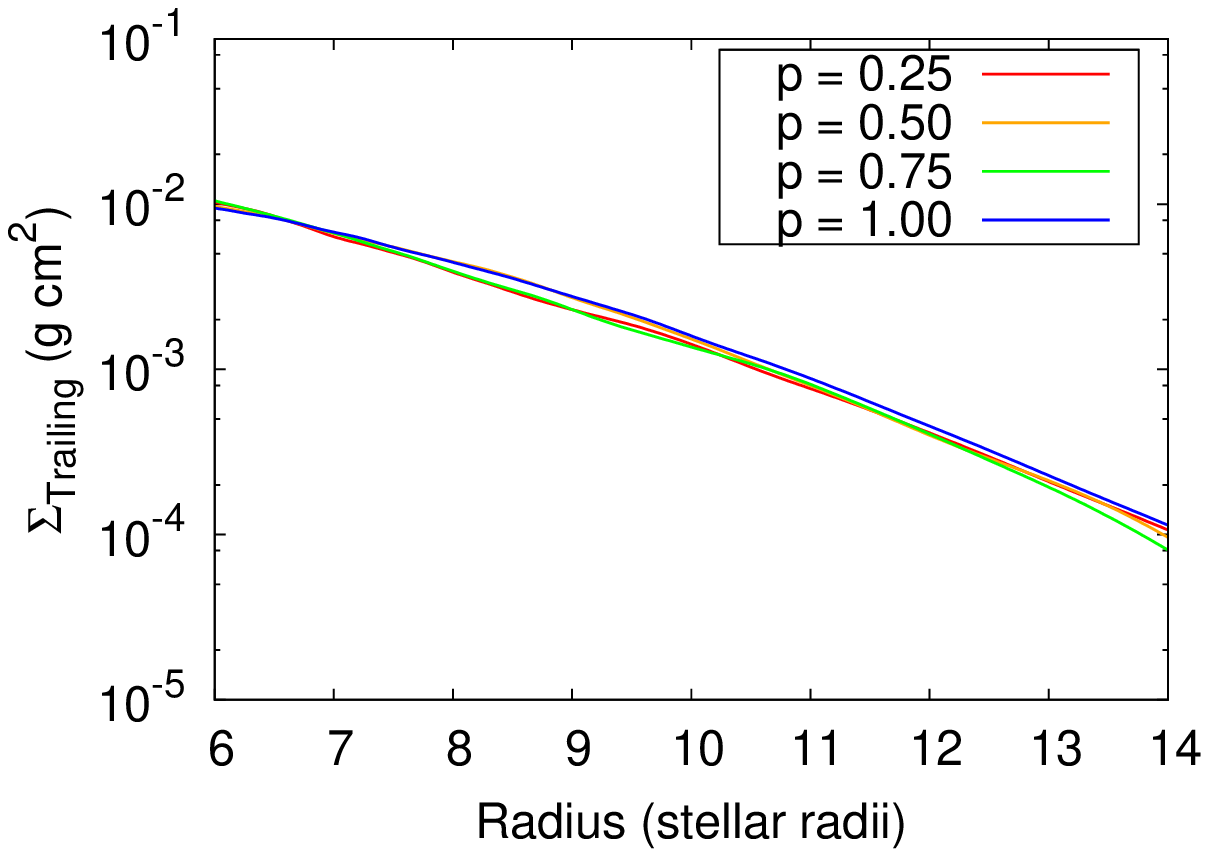}} 
\\
\subfigure[$\theta = 45\degr$]
{\includegraphics[width=0.28\textheight,keepaspectratio]{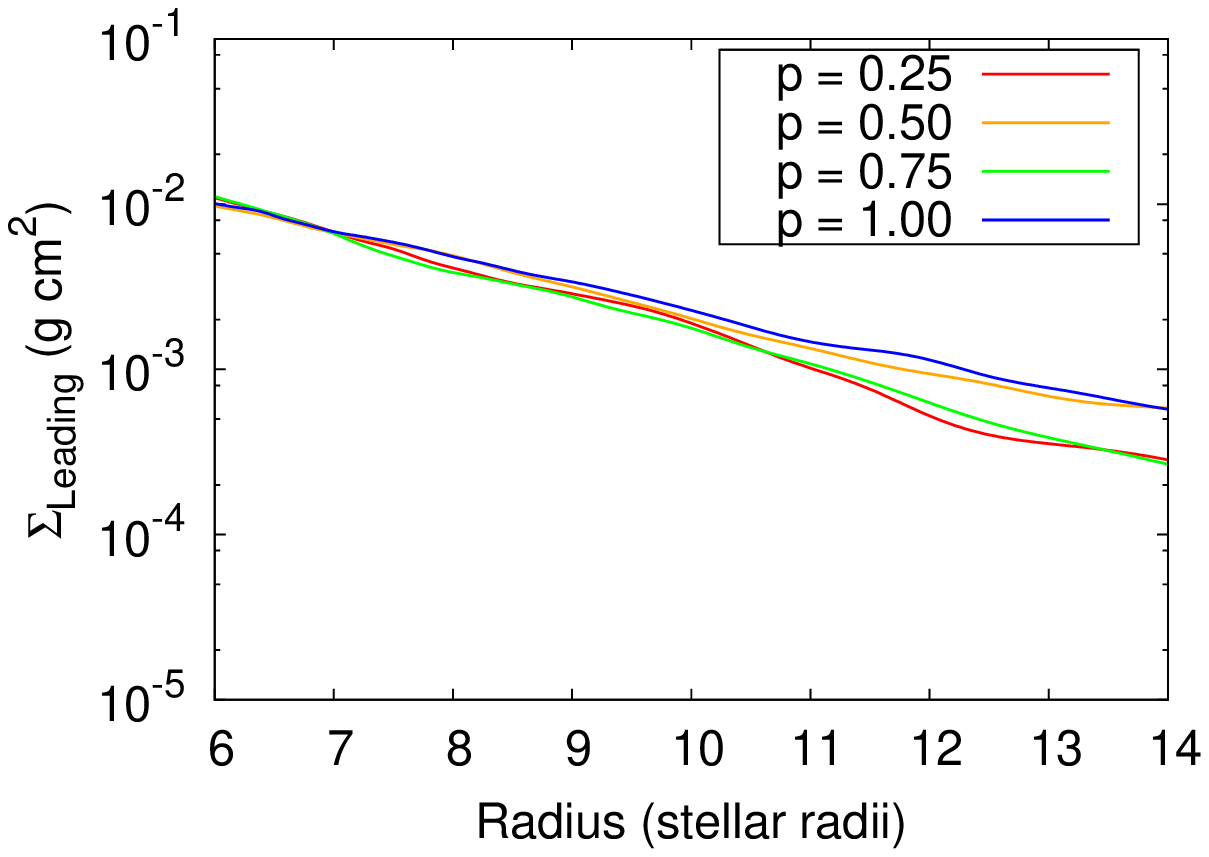}} 
\subfigure[$\theta = 45\degr$]
{\includegraphics[width=0.28\textheight,keepaspectratio]{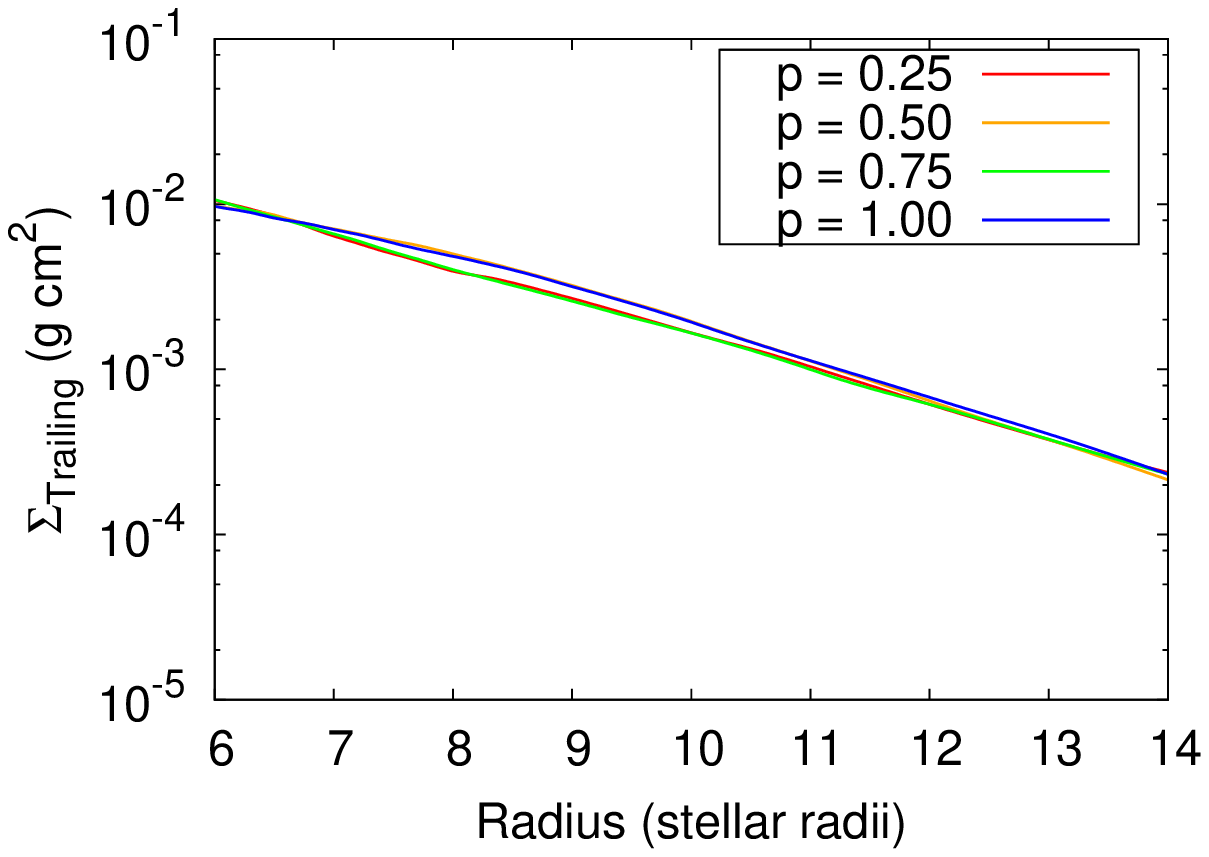}} 
\\
\subfigure[$\theta = 60\degr$]
{\includegraphics[width=0.28\textheight,keepaspectratio]{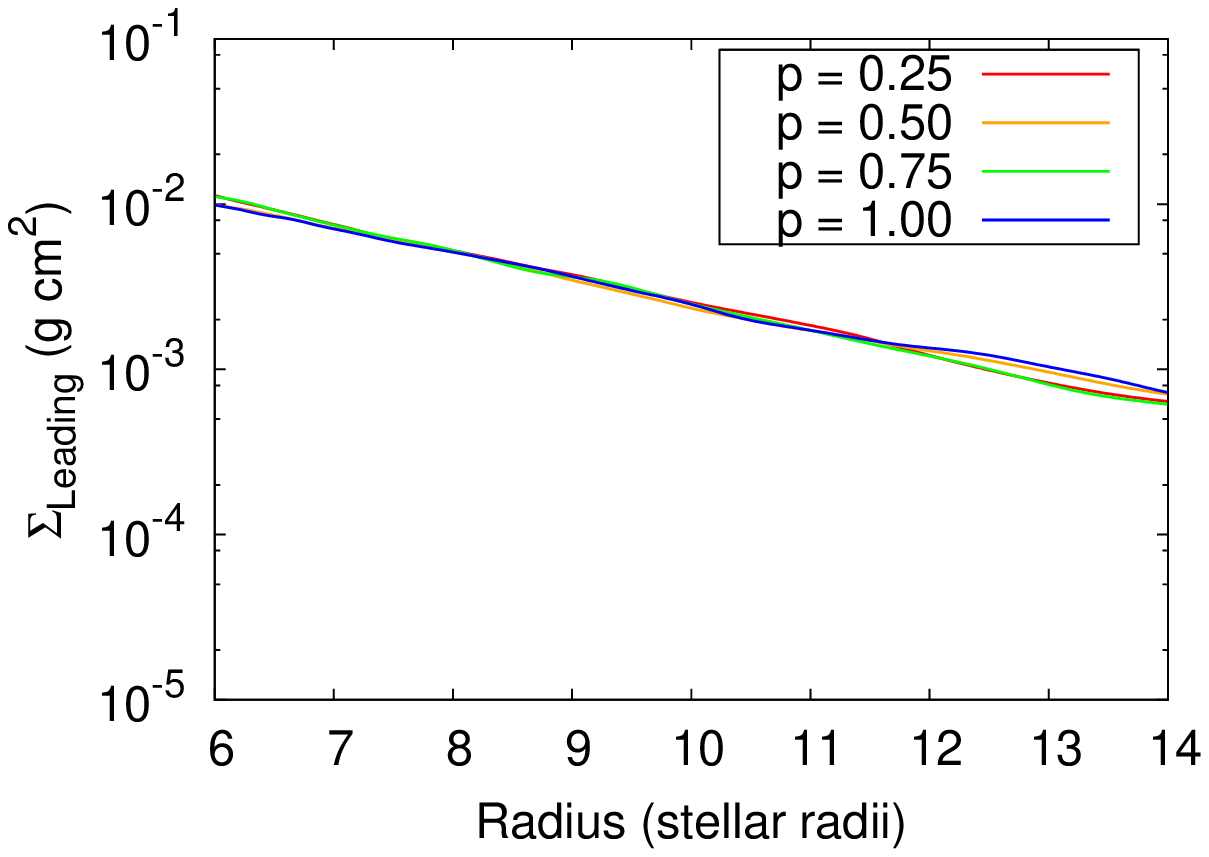}}
\subfigure[$\theta = 60\degr$]
{\includegraphics[width=0.28\textheight,keepaspectratio]{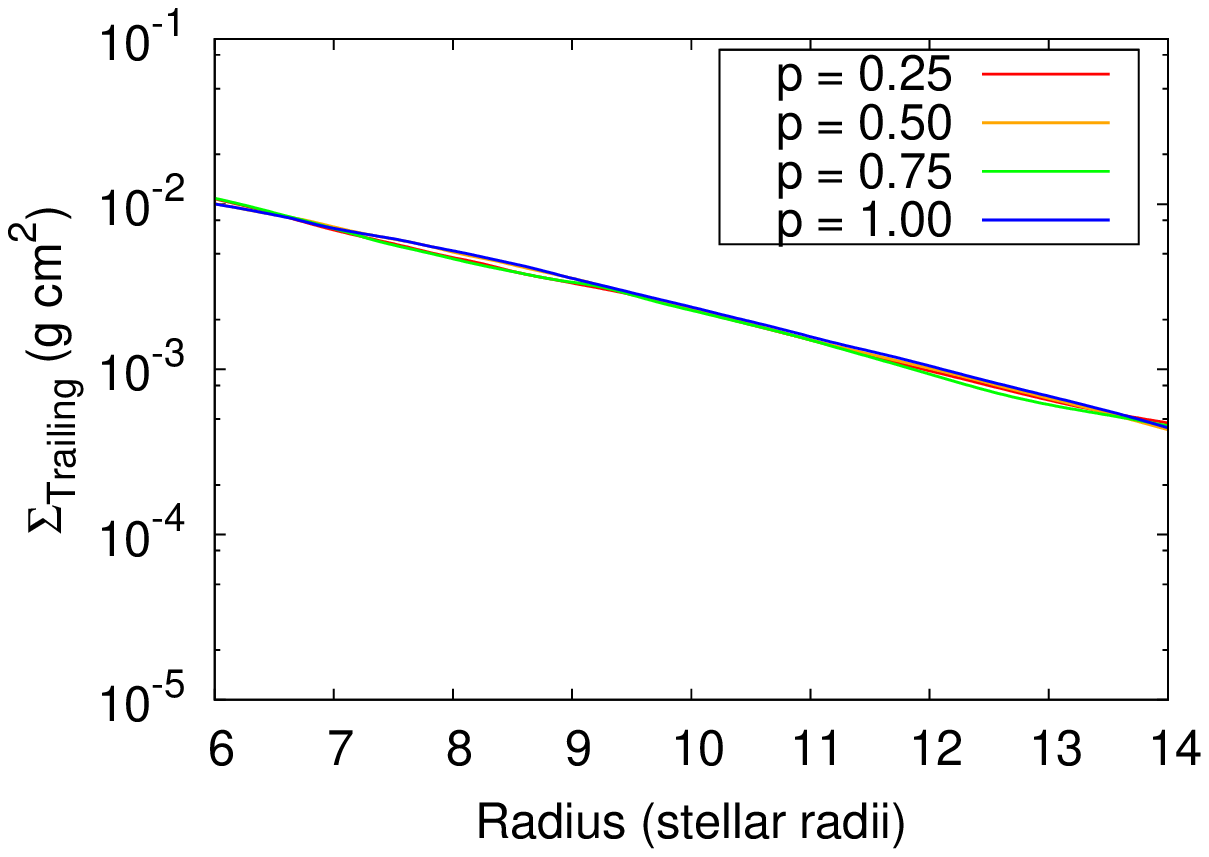}} 
\caption[Phase dependence of surface densities in leading arms]{Comparison of the surface density of the arms as a function of radial distance, $r$, at different orbital phases; $p$ = 0.25 (red), 0.50 (orange), 0.75 (green), and 1.00 (blue) with the leading arms and trailing arms presented on the left and right panels, respectively. From top to bottom, the panels show the results for misalignment angles of $\theta$ = 0$\degr$, 30$\degr$, 45$\degr$, and 60$\degr$, respectively. The viscosity of all models presented is $\alpha_\mathrm{SS}$ = 0.5.}
\label{fig4:peak_ph_l}
\end{figure*}

Using Equation~\ref{eq4:sigma_phi} we can study the density of the spiral arms in greater detail. Figure~\ref{fig4:peak_ph_l} compares the surface density fall-off along the leading and trailing arms, $\Sigma_{arm}$, respectively, as a function of $r$ for different phases of the binary orbit, all for a viscosity parameter of $\alpha_\mathrm{SS}$ = 0.5. We see that the phase has very little effect on the surface density of the leading arm in both the $\theta$ = 0$\degr$ (aligned) and $\theta$ = 60$\degr$ systems. However, we do see variation in the outer disk of the $\theta$ = 30$\degr$ and $\theta$ = 45$\degr$ systems. In both instances, the surface density of the arm is greater at phases of $p$ = 0.5 and 1.0, when the secondary intersects the plane of the disk in the misaligned case. In all four cases, the surface density of the trailing arm is independent of phase since it does not seem to be as directly affected by the secondary and therefore, not directly  dependent on the vertical position of the secondary during its orbit. We note that Figure~\ref{fig4:peak_ph_l} provides yet another verification that the structures of the arms are phase-locked with the secondary's motion.

\begin{figure*}
\center

\subfigure[$\theta = 0\degr$]
{\includegraphics[width=0.35\textwidth,keepaspectratio]{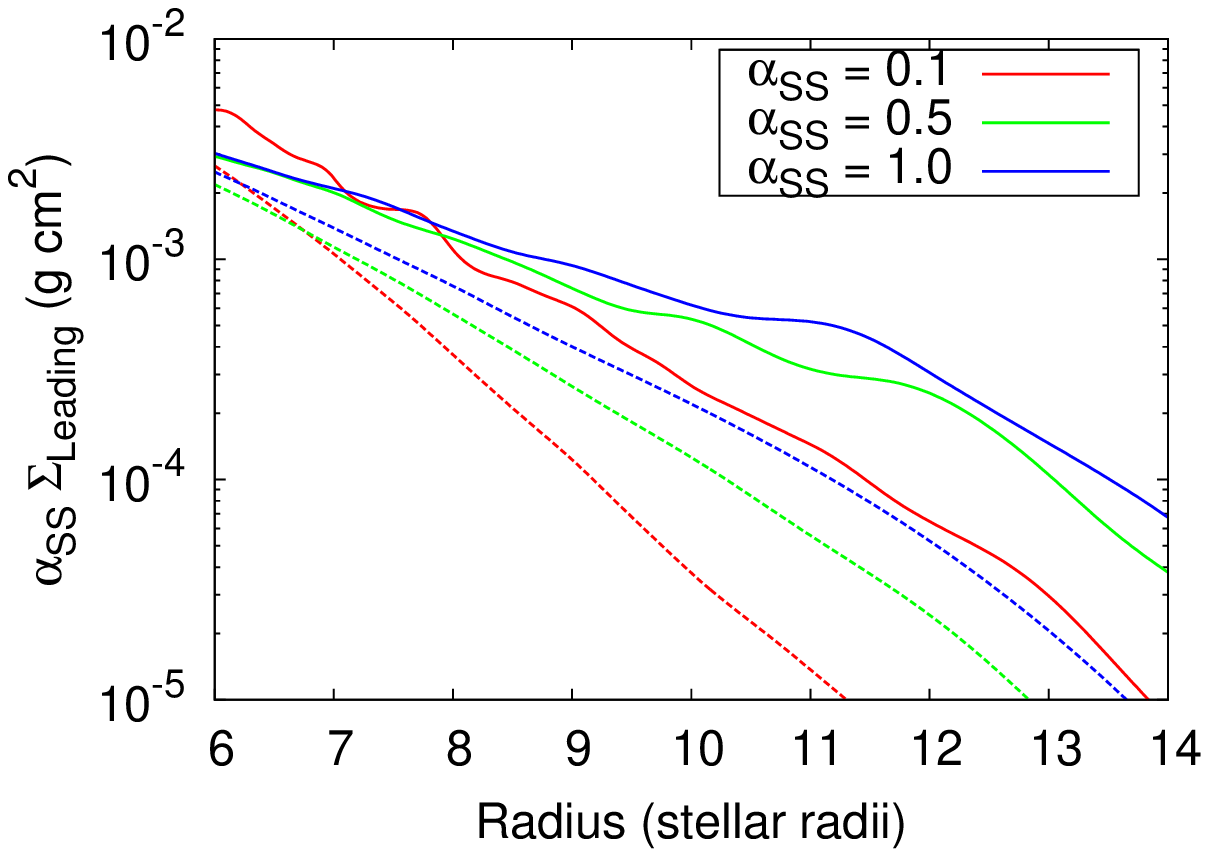}} 
\subfigure[$\theta = 0\degr$]
{\includegraphics[width=0.35\textwidth,keepaspectratio]{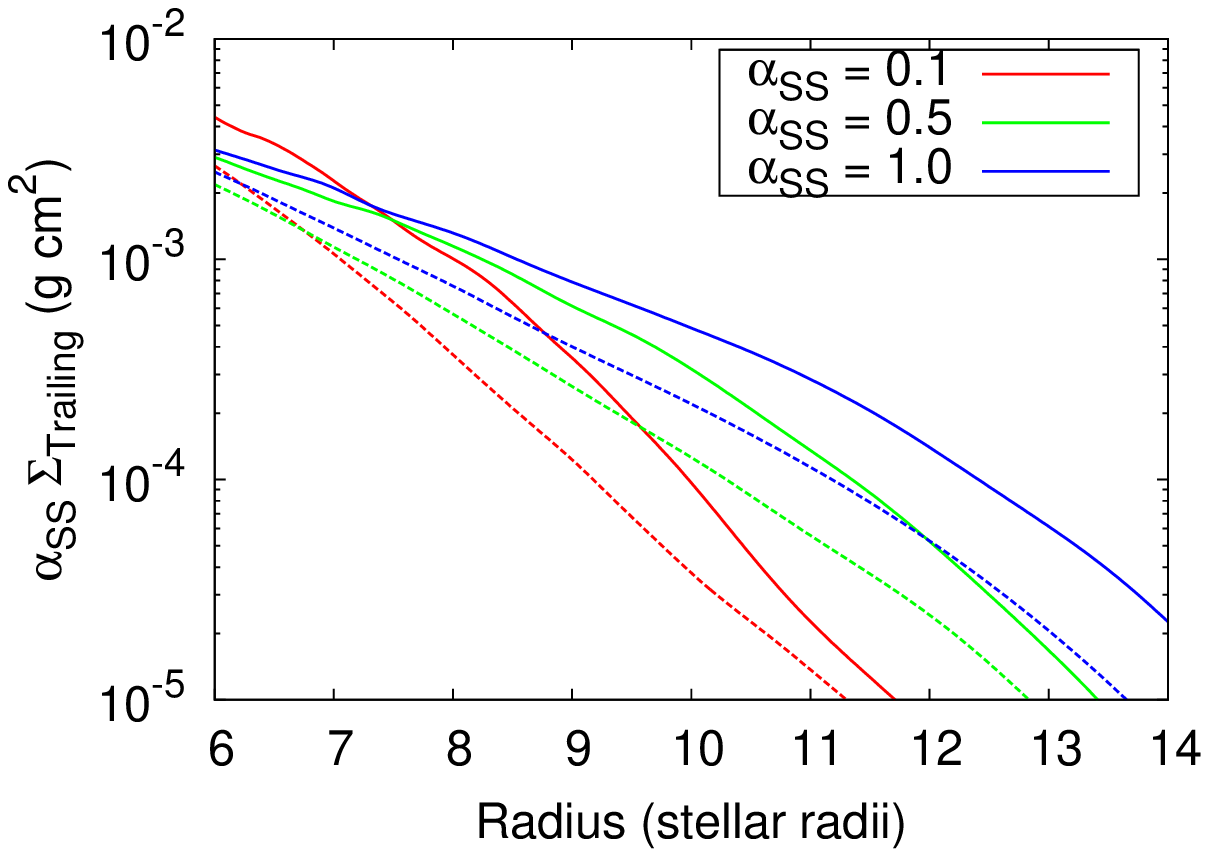}}
\\
\subfigure[$\theta = 30\degr$]
{\includegraphics[width=0.35\textwidth,keepaspectratio]{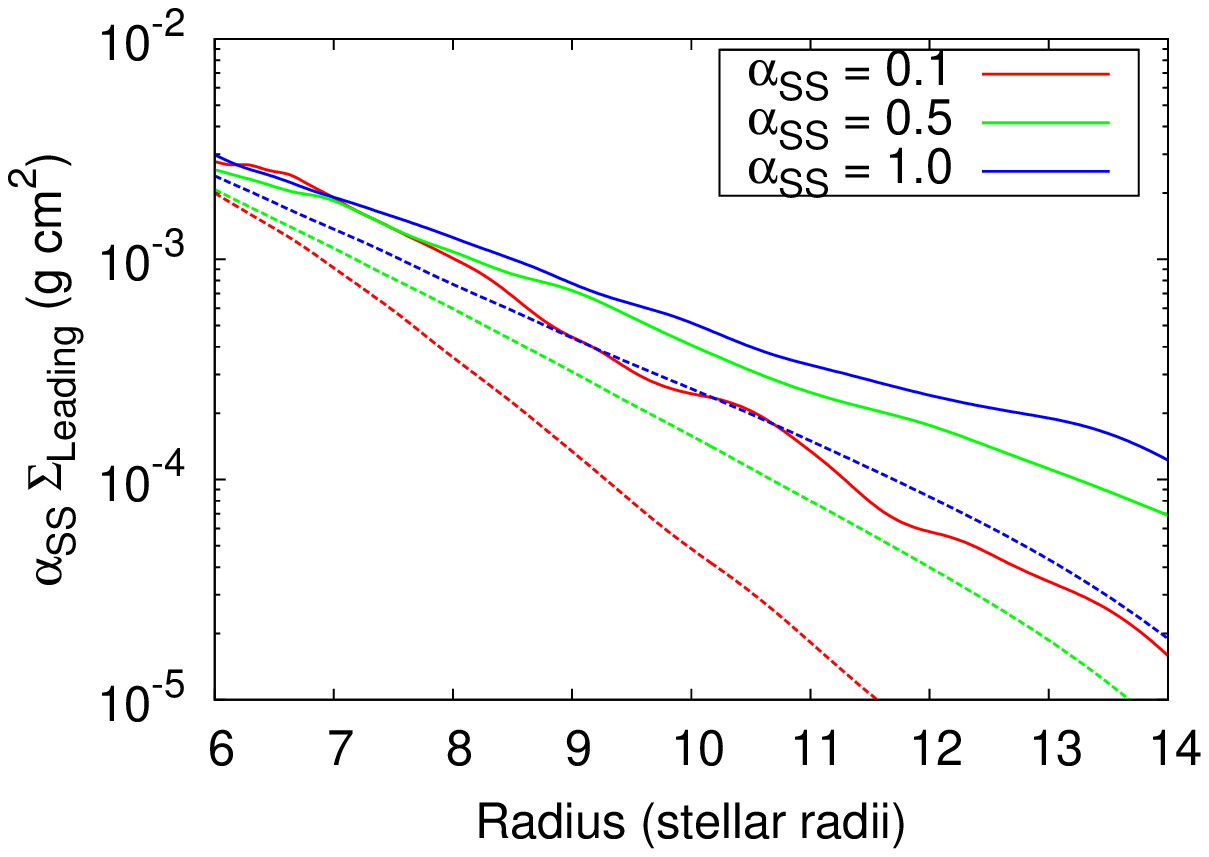}} 
\subfigure[$\theta = 30\degr$]
{\includegraphics[width=0.35\textwidth,keepaspectratio]{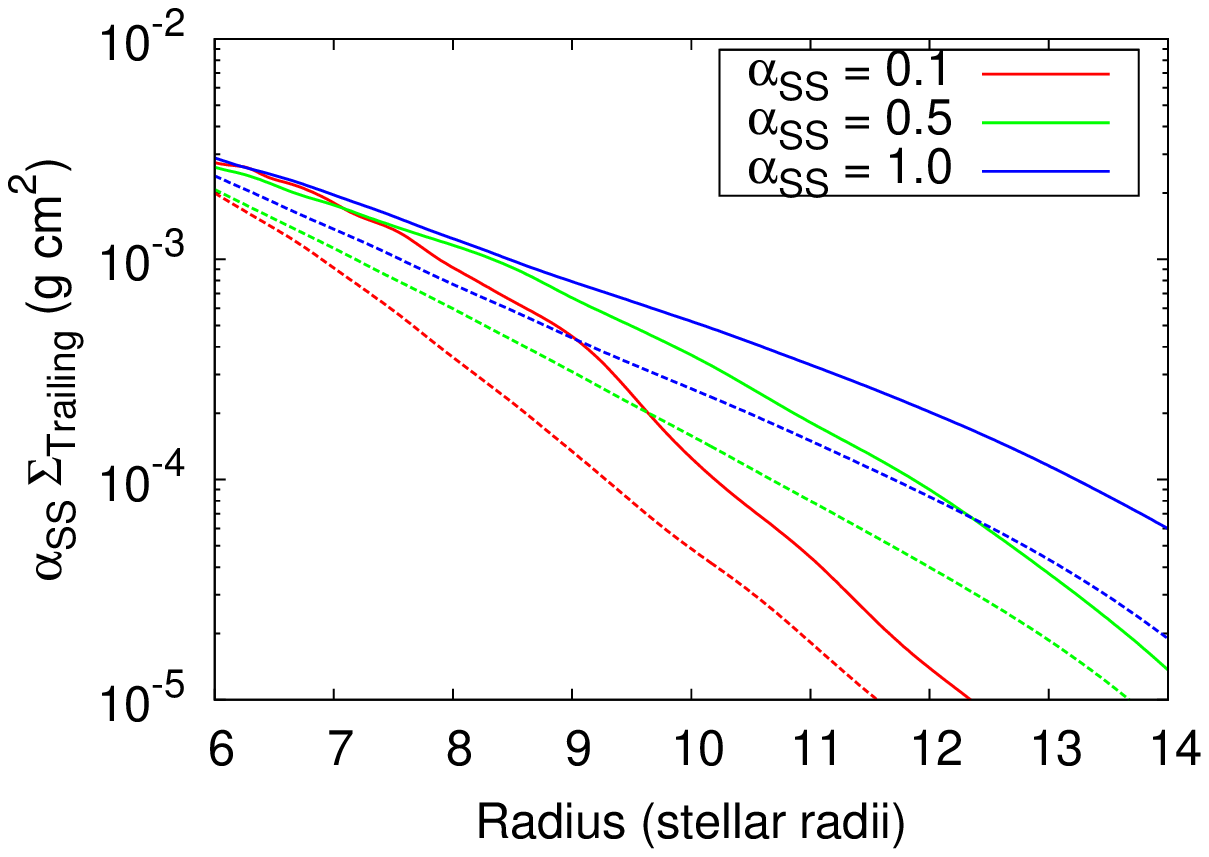}}
\\
\subfigure[$\theta = 45\degr$]
{\includegraphics[width=0.35\textwidth,keepaspectratio]{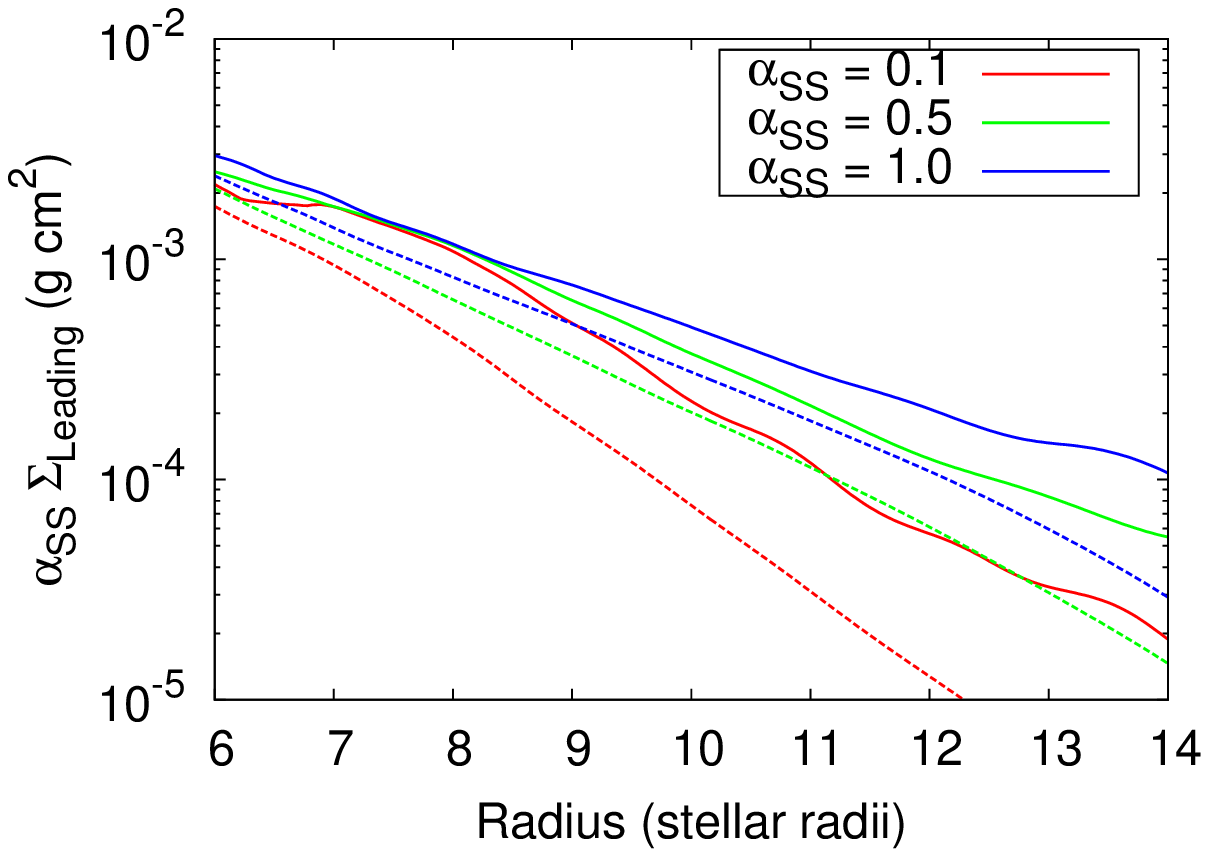}} 
\subfigure[$\theta = 45\degr$]
{\includegraphics[width=0.35\textwidth,keepaspectratio]{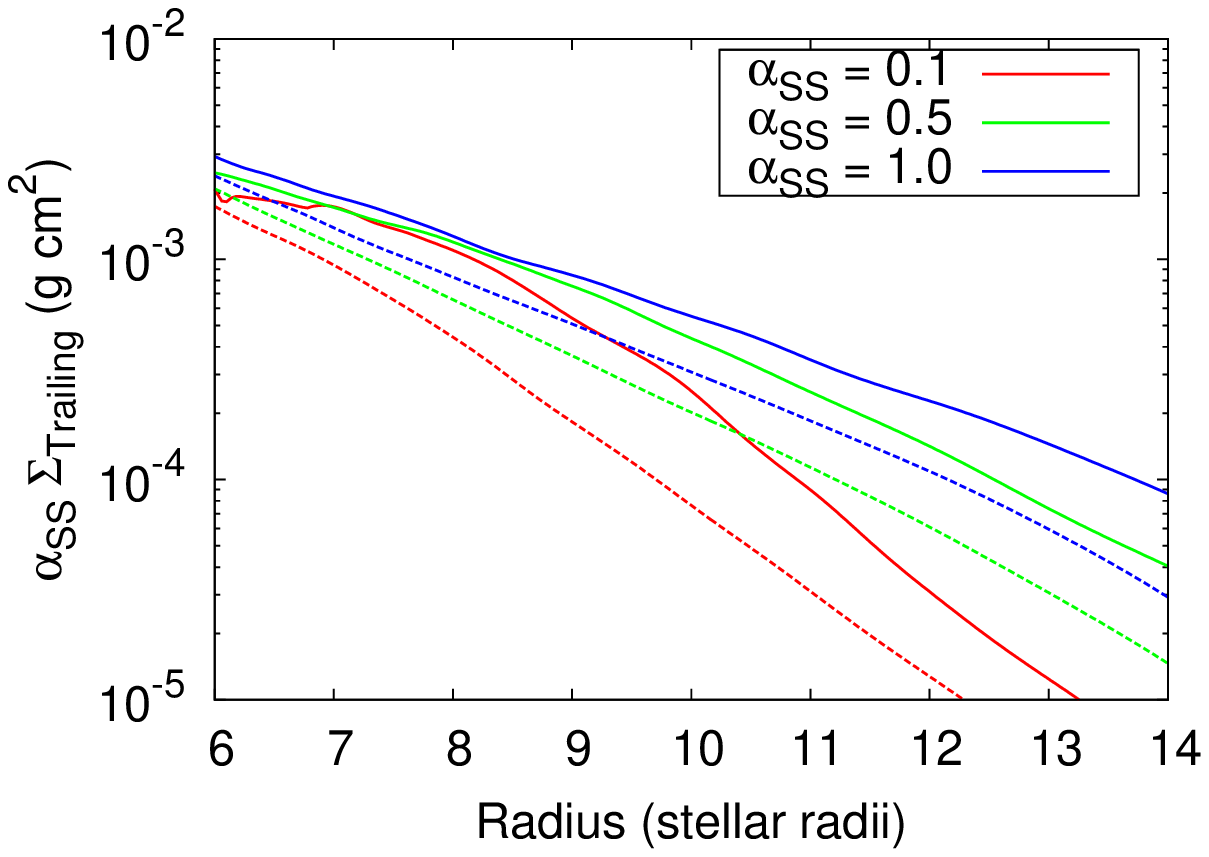}} 
\\
\subfigure[$\theta = 60\degr$]
{\includegraphics[width=0.35\textwidth,keepaspectratio]{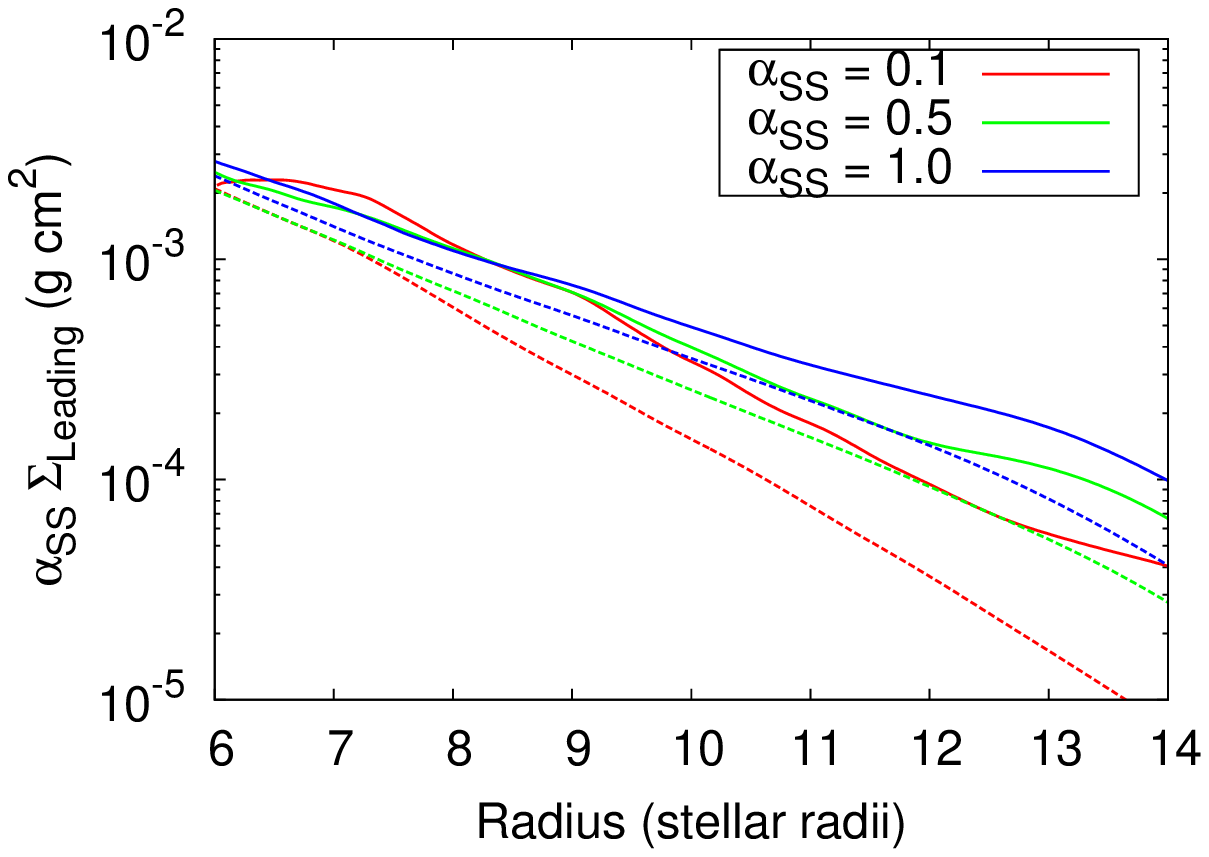}} 
\subfigure[$\theta = 60\degr$]
{\includegraphics[width=0.35\textwidth,keepaspectratio]{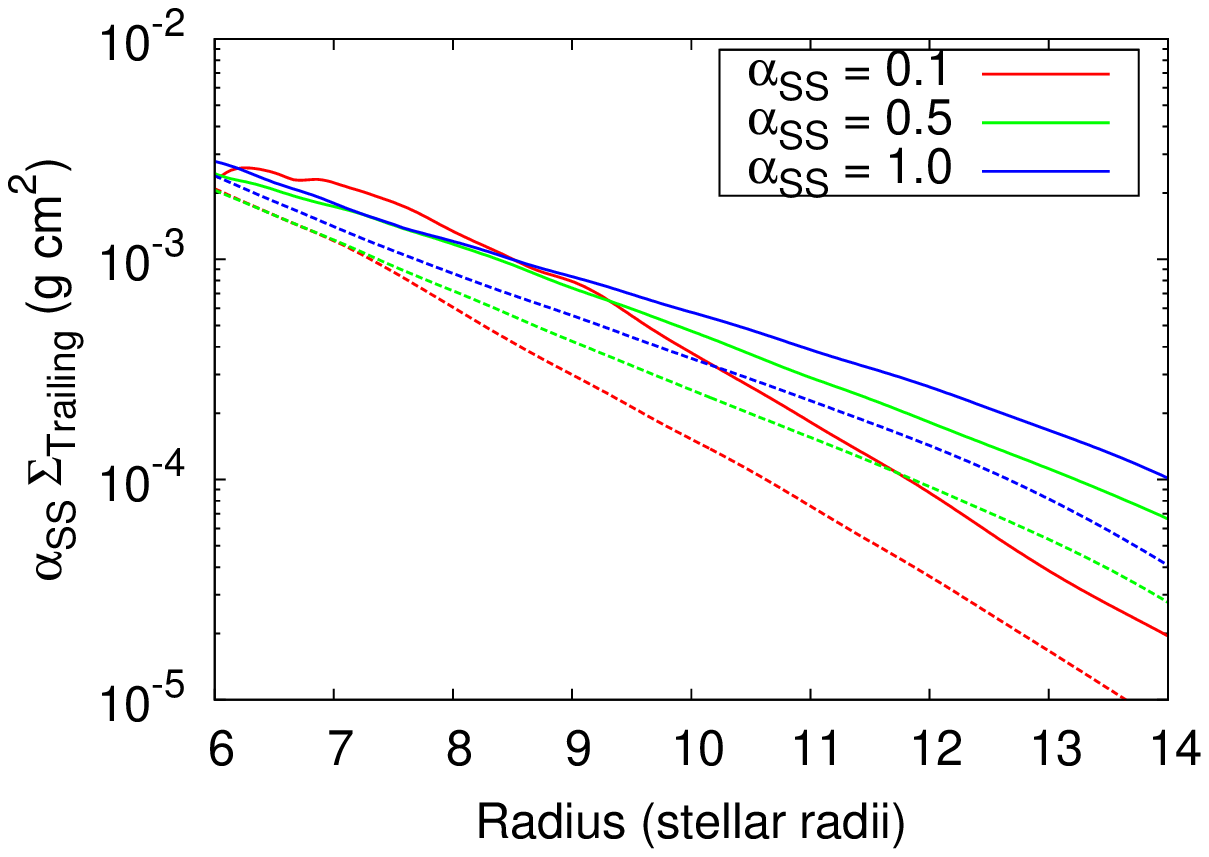}} 

\caption[Viscosity dependence of surface densities in leading arms]{Similar to Figure~\ref{fig4:peak_ph_l} for different viscosity values; $\alpha_\mathrm{SS}$ = 0.1 (red), 0.5 (green), and 1.0 (blue). The solid lines show the surface densities of the density enhancements while the dashed lines shows the azimuthal-averaged surface densities of that disk. The surface densities were multiplied with $\alpha_\mathrm{SS}$ and the timescale of each run is also adjusted to account for the different viscous timescales. The curve for $\alpha_\mathrm{SS}$ = 0.1 is shown at 50 orbital periods, $\alpha_\mathrm{SS}$ = 0.5 at 10 orbital periods, and $\alpha_\mathrm{SS}$ = 1.0 at 5 orbital periods so that these are at the same evolutionary state. The orbital phase of all models presented is $p$ = 1.00. }
\label{fig4:peak_a_l}
\end{figure*}

Figure~\ref{fig4:peak_a_l} shows a comparison of the surface density of each arm (solid lines) for all three viscosity parameters, $\alpha_\mathrm{SS}$ = 0.1 (red), $\alpha_\mathrm{SS}$ = 0.5 (green), and $\alpha_\mathrm{SS}$ = 1.0 (blue), with each row in representing a different misalignment angle. As \citet{bjo05} have shown, the surface density of the disk is expected to scale with $\alpha_\mathrm{SS}^{-1}$, therefore we multiplied each profile by $\alpha_\mathrm{SS}$ for our comparison. Furthermore, given that the diffusion timescale of the disk is directly related to the viscosity of the disk, the time for each snapshot for each value of $\alpha_\mathrm{SS}$ was changed so that our comparison is at the same evolutionary time. Therefore Figure~\ref{fig4:peak_a_l} shows the results of the $\alpha_\mathrm{SS}$ = 0.1  simulation at 50 orbital periods, $\alpha_\mathrm{SS}$ = 0.5 at 10 orbital periods, and $\alpha_\mathrm{SS}$ = 1.0 at 5 orbital periods. Finally the dashed lines represent the azimuthal-averaged surface density of the disk as a means of comparison. All panels are at a phase of 1.00.

Here we see that viscosity has a visible effect on the fall-off rate of the surface density profiles, showing a steeper drop in density for smaller $\alpha_\mathrm{SS}$ values. We also notice that for the more viscous disks ($\alpha_{SS}$ = 0.5 and 1.0), the surface density profiles of the density enhancements (solid lines) have similar characteristics of the azimuthal-averaged density profiles (dashed lines). This feature is more noticeable in the trailing arms and models with a higher misalignment angle. Differences between the two density profiles are more pronounced in low viscosity models ($\alpha_{SS}$ = 0.1), especially for the models with low misalignment angles. These differences can be explained by the fact that disks with lower viscosities, are more affected by the passing of the secondary. Finally, we also see that for low viscosity disk, the fall-off rate of the surface density profile is much lower in the inner part of the disk than the outer part. This is particularly evident in the trailing arm but is also present in the leading arm. This is due to the accumulation affect, i.e., the build-up of density inside the truncation radius, as previously noted by other researchers and as discussed in the introduction. 

To better understand the winding characteristics in these spiral features it is important to recall that the arms are formed because the individual orbits of the particles or packets of gas in the disk are not circular, but rather elliptical, owing to the tidal perturbations of the secondary. The arms trace the positions of the apastron of the orbits, the higher densities being simply the result of lower orbital speeds at apastron.  The larger the inclination of the orbit, the smaller the tidal perturbations on the disk and therefore the less eccentric the orbits. As a result, the spirals become more tightly wound. Conversely, for the coplanar model ($\theta = 0\degr$), $\gamma$ has the largest values, indicating spiral arms that spread further out in the disk. The dependence of $\gamma$ on $\alpha_\mathrm{SS}$, on the other hand, is easily understood considering that viscosity acts towards circularizing the orbits of the particles, and larger values of $\alpha_\mathrm{SS}$ imply a stronger viscous coupling of the gas.

\begin{figure*}
\center
\subfigure[$\alpha_\mathrm{SS} = 0.1$, leading-arm]
{\includegraphics[width=0.4\textwidth,keepaspectratio]{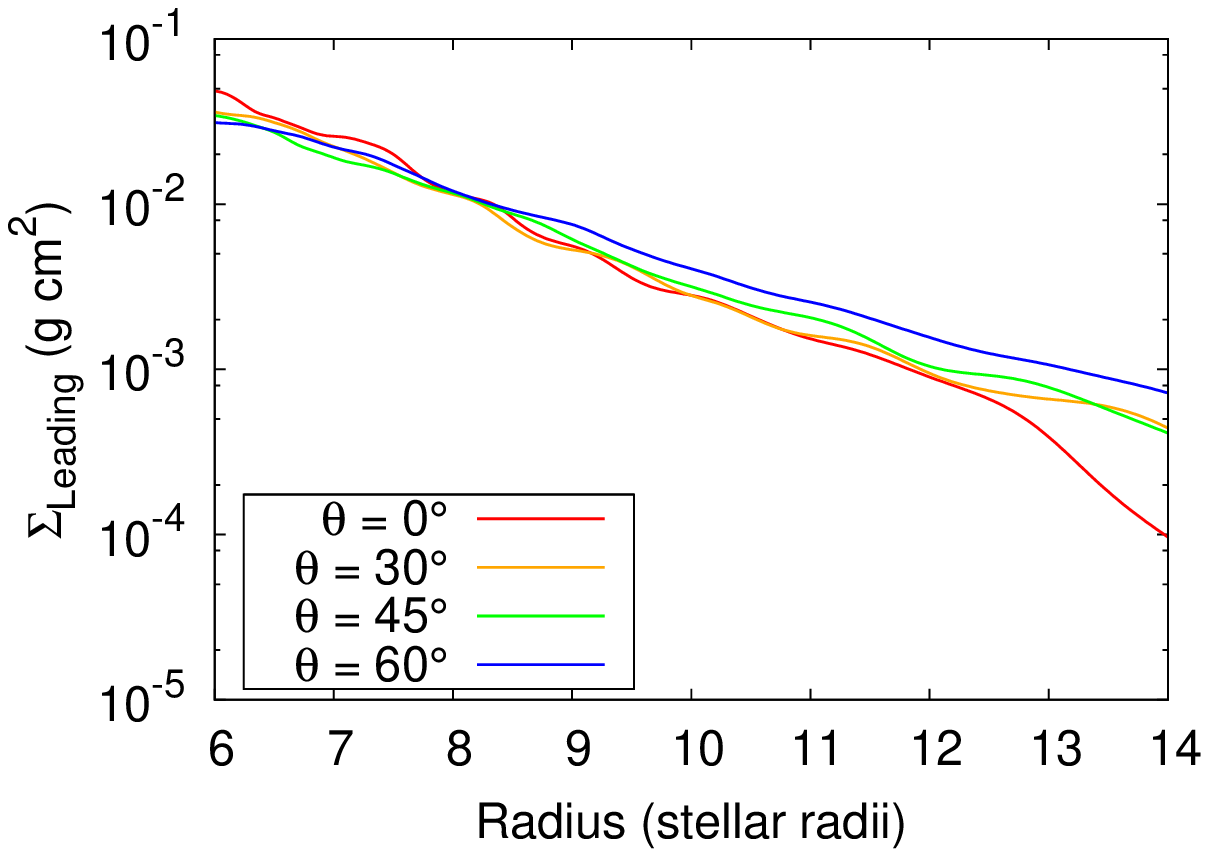}} 
\subfigure[$\alpha_\mathrm{SS} = 0.1$, trailing-arm]
{\includegraphics[width=0.4\textwidth,keepaspectratio]{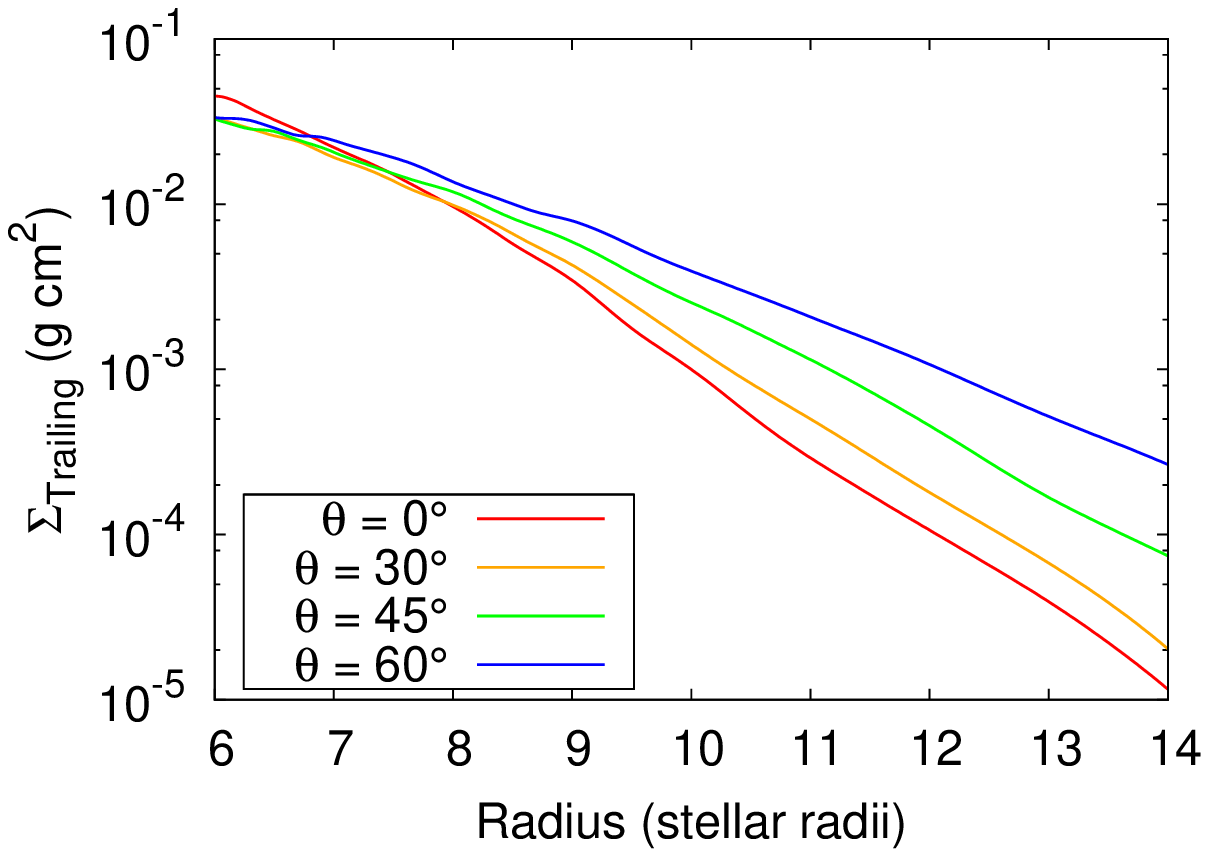}} \\
\subfigure[$\alpha_\mathrm{SS} = 0.5$, leading-arm]
{\includegraphics[width=0.4\textwidth,keepaspectratio]{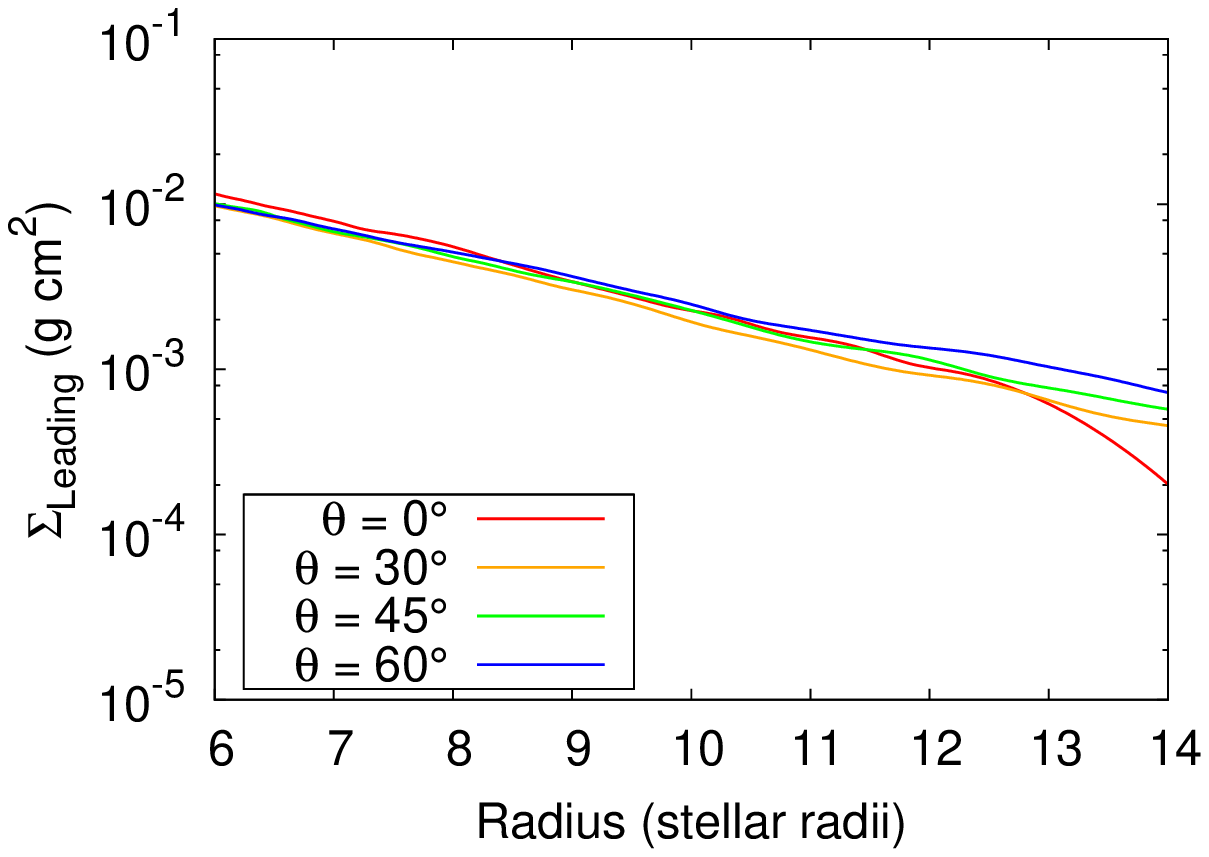}} 
\subfigure[$\alpha_\mathrm{SS} = 0.5$, trailing-arm]
{\includegraphics[width=0.4\textwidth,keepaspectratio]{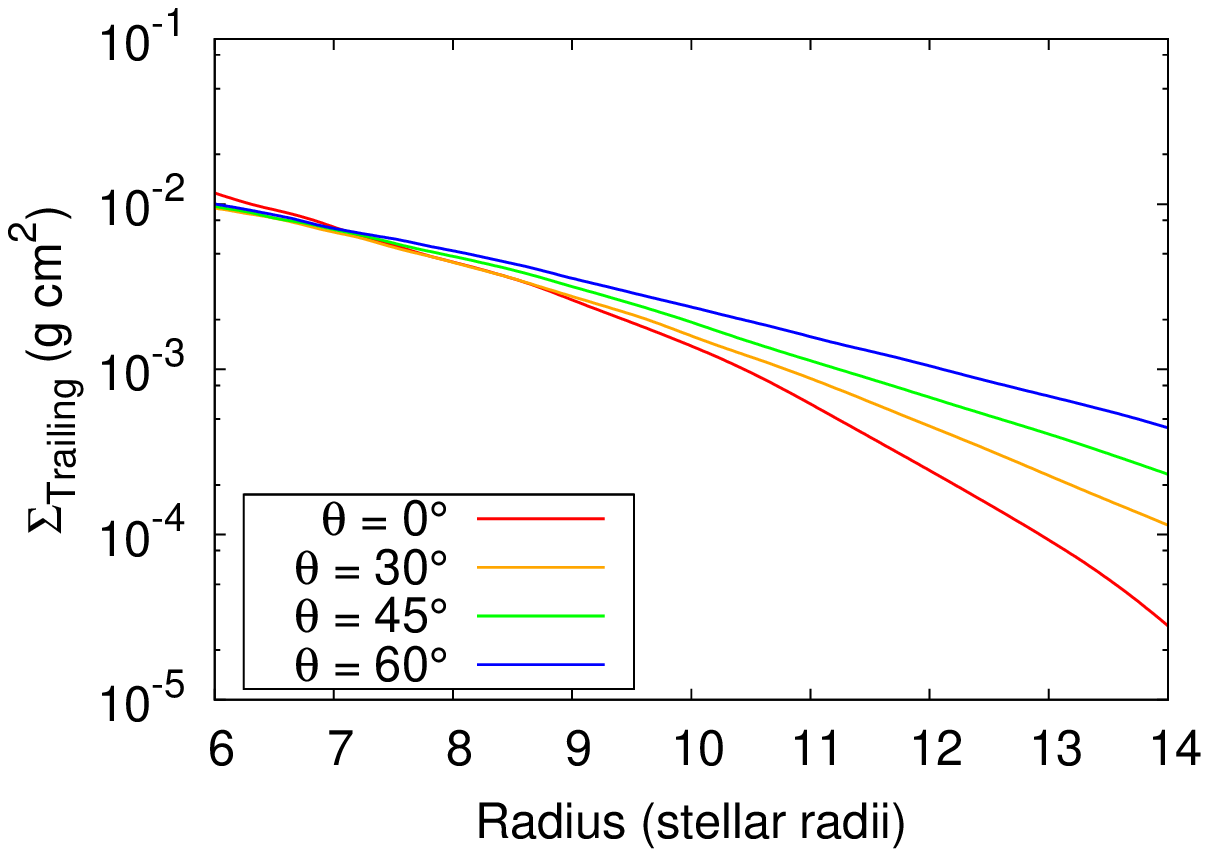}} \\
\subfigure[$\alpha_\mathrm{SS} = 1.0$, leading-arm]
{\includegraphics[width=0.4\textwidth,keepaspectratio]{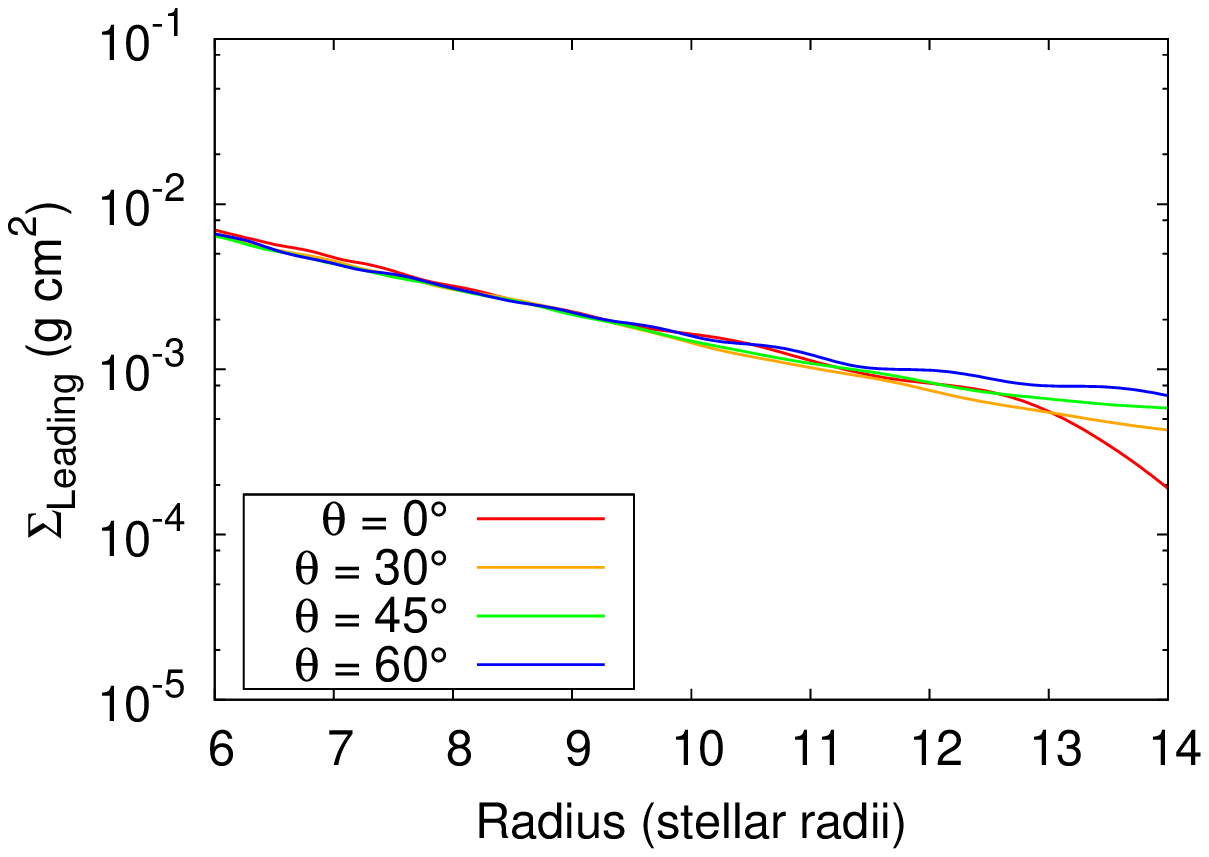}} 
\subfigure[$\alpha_\mathrm{SS} = 1.0$, trailing-arm]
{\includegraphics[width=0.4\textwidth,keepaspectratio]{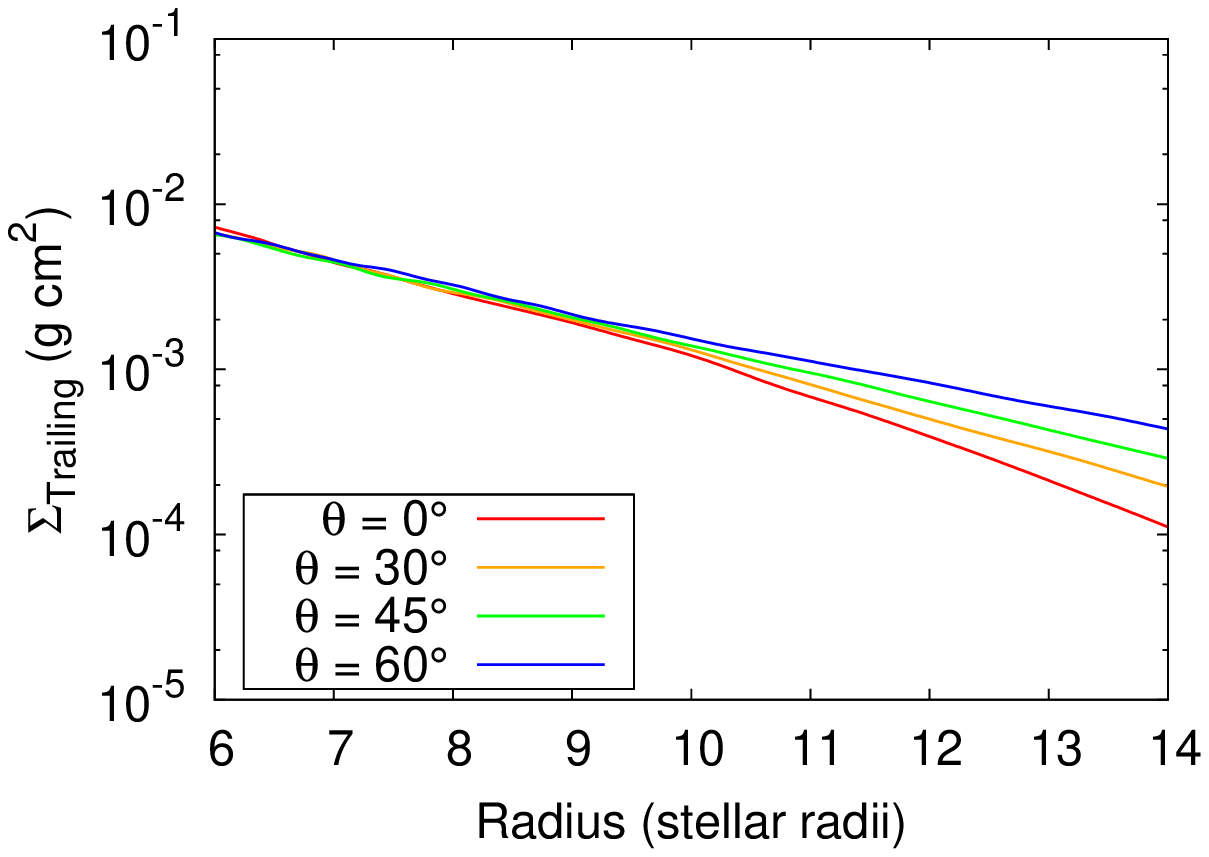}} \\
\caption[Misalignment angle dependence of spiral arms surface densities.]{Comparison of the surface density of the leading (left panels) and trailing arms (right panels) as a function of radial distance, $r$, for different misalignment angles; $\theta$ = 0$\degr$ (red), 30$\degr$ (orange), 45$\degr$ (green), and 60$\degr$ (blue). From top to bottom, the panels show the results for viscosity values of $\alpha_\mathrm{SS}$ = 0.1, 0.5, and 1.0, respectively. The orbital phase of all models presented is $p$ = 1.00.}
\label{fig4:peak_th}
\end{figure*}

Figure~\ref{fig4:peak_th} compares the surface density variations of each arm for all four misalignment angles, $\theta$ = 0$\degr$ (red), 30$\degr$ (orange), 45$\degr$ (green), and 60$\degr$ (blue), in a similar fashion as Figures~\ref{fig4:peak_ph_l} and \ref{fig4:peak_a_l}. This time, the rows, from top to bottom, represent viscosity parameters of $\alpha_\mathrm{SS}$ = 0.1, 0.5, and 1.0, respectively. Once again, the phase in all six panels is 1.00. Here we see that variations are very small close to the star but increase with radial distance, with surface densities decreasing with $r$ more slowly when the system is increasingly misaligned. We notice that the location where the curves start to diverge, increases with the angle of misalignment. We also notice that the location of this divergence of the trailing arms is always closer to the star compared with the leading arm.

\subsection{Interferometric Predictions}

\begin{figure*}
\center
\includegraphics[width=0.45\textwidth]{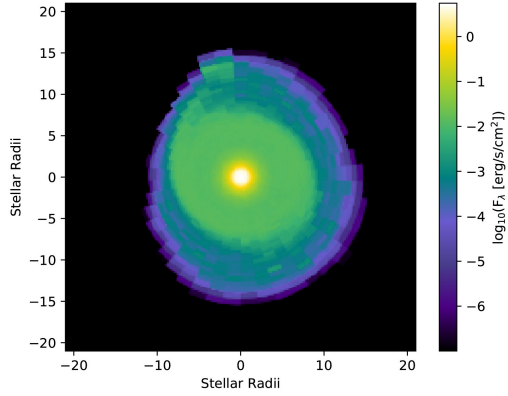}
\caption[Halpha image]{The H$\alpha$ emitting region covering the wavelength range from 6551.5 to 6576.5 \AA \: from our co-planar SPH model using \textsc{hdust} to compute the flux at a pole-on inclination over the region spanning 20 by 20 stellar radii. The colour scale corresponds to the flux with the key is provided to the right of the image.}
\label{hdusthalpha}
\end{figure*}

\begin{figure*}
\center
\includegraphics[width=0.45\textwidth]{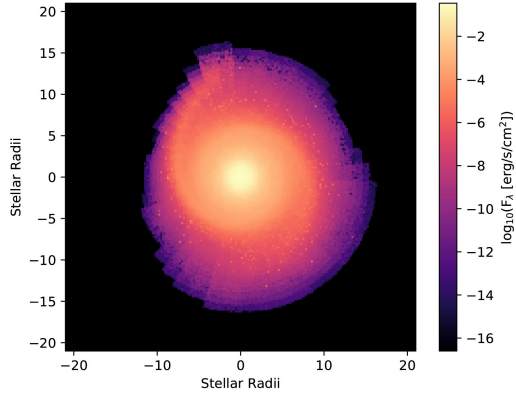}
\caption[K-band image]{The same as Figure~\ref{hdusthalpha} except for the K-band from 2.0 to 2.4 microns.}
\label{hdustkband}
\end{figure*}

For simplicity, we decided to focus our interferometry analysis of our SPH models for the case where the companion is co-planar and for the phase, $p = 1.0$. Figure~\ref{hdusthalpha} shows the pole-on view of the H$\alpha$-emitting region that was obtained by integrating the flux values produced by \textsc{hdust} over the wavelength region from 6551.5 to 6576.5 \AA \:. Similarly, Figure~\ref{hdustkband} shows the K-band image that corresponds to flux values integrated over the wavelength range from 2.0 to 2.4 $\mu$m. We decided to present both Figures~\ref{hdusthalpha} and~\ref{hdustkband} out to 20 stellar radii in x and y for ease of comparison with the surface density plots presented in Figure~\ref{spiral and phases}. The spiral density enhancements are more predominant in the K-band~(Figure~\ref{hdustkband}) due to greater contrast since at H$\alpha$ emitting wavelengths the disk is optically thick and essentially acts as a pseudo-photosphere with nearly constant emissivity \citep{vie15}. We also note that the H$\alpha$-emitting region (Figure~\ref{hdusthalpha}) does appear to be more radially extended compared to the K-band region, however, in model images such as those shown in Figures~\ref{hdusthalpha} and~\ref{hdustkband} this does depend on the choice of the contour levels.

Following the procedure outlined in Section~\ref{method}, the 2D discrete Fourier transforms were computed for the H$\alpha$ and K-band wavelength model images. To obtain interferometric signatures that would correspond to a range of baseline orientations, the cross-sections of the Fourier-transformed image were taken at one degree intervals (measured as a principal angle from the spatial frequency axis that corresponds to the x-axis in the image plane), all of which pass through the origin of the Fourier space. The normalized Fourier power (i.e., the real and imaginary parts of the transform added together in quadrature) at each principal angle (in steps of 1$\degr$) was then fit with a Gaussian function to obtain the corresponding FWHM measure, and along with a similar measure in orthogonal direction allowed us to obtain an axis ratio directly from the ratio of these FWHM pairs. 

\begin{figure*}
\center
\includegraphics[width=0.45\textwidth]{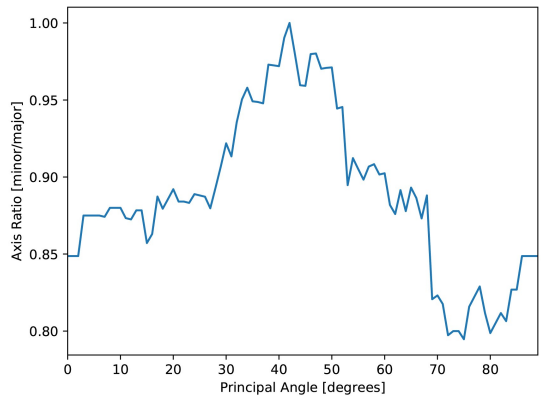}
\caption[]{The axis ratio obtained from Gaussian fits at two orthogonal cross-sections of Fourier transformed H$\alpha$ image, with one of the cross-sections being oriented at a principle angle measured from the spatial frequency axis that corresponds to the x-direction in the image space (repeated in steps of 1$\degr$ for different principal angles).}
\label{halpharatios}
\end{figure*}

\begin{figure*}
\center
\includegraphics[width=0.45\textwidth]{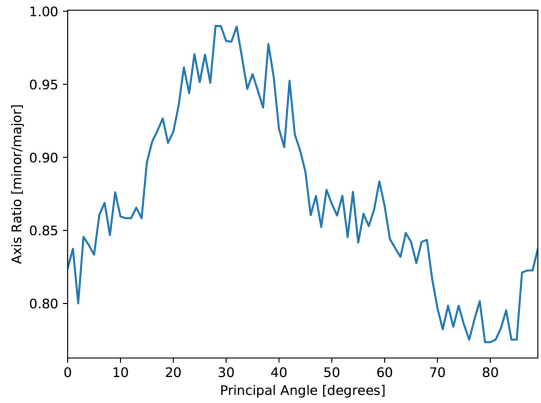}
\caption[]{Same as Figure~\ref{halpharatios}, but for the K-band.}
\label{kbandratios}
\end{figure*}

The effective axis ratios for the H$\alpha$ and K-band emitting regions are shown in Figures~\ref{halpharatios} and~\ref{kbandratios}. We found that the ratio could be as small as 0.79, compared to the ratio of 1.0 that would be expected for pole-on circular disk in H${\alpha}$. For the K-band we found for this particular model that the axis ratio could be as small as 0.77. This means that when density enhancements are present in the disk, depending on the orientation of interferometric baselines, the density enhancements may cause a substantial departure from circular symmetry even for systems that are viewed very close to pole-on. Finally, we note that the ratios of orthogonal disk cross sections plotted in Figures~\ref{halpharatios} and~\ref{kbandratios} are quite variable and this is because the extent of the disk for each cross section was determined at discrete intervals and reflects the complex nature of density enhancements within the disk. 

Although, our modeling of the expected interferometric signatures was focused on the disk model that was viewed pole-on, we also investigated the expected axis ratios for a range of other inclinations, following the same procedure as described above (using a model that corresponds to the same phase of $p=1$). We note that with increasing viewing angle, other effects such as disk opacity and re-absorption of light within the disk become increasingly important as the the inclination angles approach an edge-on arrangement with respect to the observer. Of course, the disk opacity will depend on the actual disk density and available ionizing radiation from the central star, which in turn depends on the stellar parameters. Although a detailed investigation of interferometric observables for a range of disk densities and stellar parameters is beyond the scope of this work, it is nevertheless worthwhile to illustrate how the axis ratios corresponding to different inclinations can be affected by the presence of the density variations within the disk.

For the H$\alpha$-emitting region at viewing angles of 30$\degr$, 60$\degr$ and 90$\degr$ (edge-on) we obtained axis ratios of 0.71, 0.48 and 0.14, respectively. Similarly, for the K-band emitting region at viewing angles of 30$\degr$, 60$\degr$ and 90$\degr$ we obtained axis ratios of 0.72, 0.49 and 0.22, respectively. We can immediately see that the axis ratios scale approximately with cosine of the viewing angle, as would be expected for optically thick, but geometrically thin disks.  However, a simple inverse cosine analysis of these axis ratios, as is commonly conducted in long-baseline interferometry, would produce inclination angles of 45$\degr$, 61$\degr$ and 82$\degr$ (instead of the 30$\degr$, 60$\degr$ and 90$\degr$).  The deviation at the largest inclination can be attributed to the effect of disk opacity and finite scale height of the disk (i.e., at large inclinations we are sensitive to the disk thickness or flaring).  In contrast, the deviation at the expected 30$\degr$ inclination can be attributed mostly to the presence of the density variations within the disk (similarly to what we have found in the case of the pole-on system where the finite thickness of the disk is not expected to play any role in deviation from circular symmetry).

\begin{figure*}
\center
\includegraphics[width=0.45\textwidth]{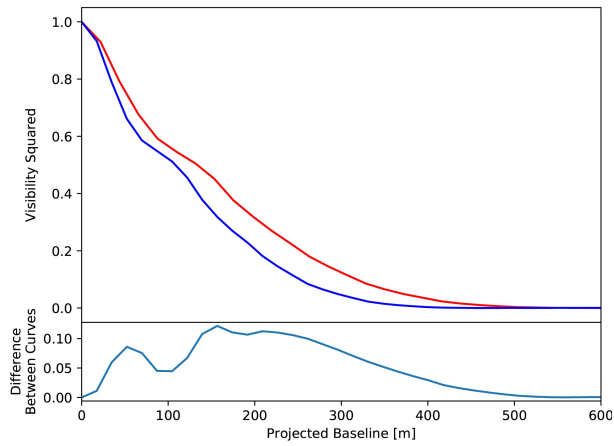}
\caption[]{The top panel shows the visibility squared versus projected baseline for the H$\alpha$ image shown in Figure~\ref{hdusthalpha}. The major and the minor axes are shown in blue and red, respectively. The difference in squared visibility corresponding to the major and minor axes is shown in the bottom panel.}
\label{vishalpha}
\end{figure*}

\begin{figure*}
\center
\includegraphics[width=0.45\textwidth]{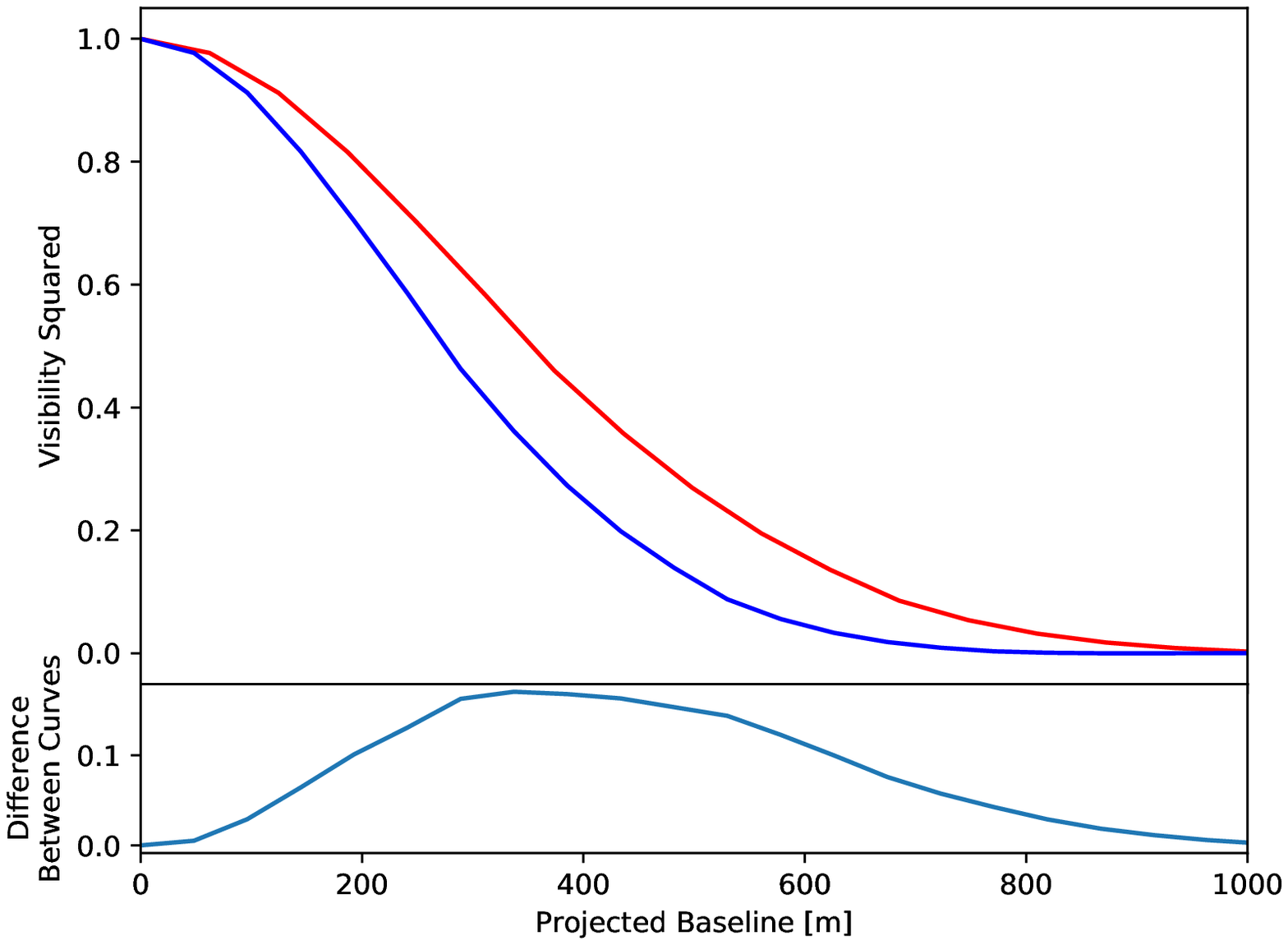}
\caption[]{Same as Figure~\ref{vishalpha}, but for the K-band image from Figure~\ref{hdustkband}.}
\label{viskband}
\end{figure*}

We also used the information contained in Figures~\ref{halpharatios} and~\ref{kbandratios} to plot visibility curves that correspond to the minor and major axes shown in the upper panels of Figures~\ref{vishalpha} and~\ref{viskband} for the H$\alpha$ and K-band emitting regions, respectively. Using the equation, $\theta = B/\lambda$ where $\theta$ is the radial spatial frequency, $B$ is the baseline length, and $\lambda$ is wavelength, we converted our spatial frequencies to projected baseline lengths in Figures~\ref{vishalpha} and~\ref{viskband}. Note that, as expected, the visibility curves corresponding to minor axes, which require longer baselines to be resolved, are always situated to the right of the curves for major axes. The lower panels of Figures~\ref{vishalpha} and ~\ref{viskband} show the difference between the interferometric signatures for major and minor axes at each projected baseline length. Also note that by comparing upper panels of Figures~\ref{vishalpha} and~\ref{viskband} we can see that the H$\alpha$ emitting region is more extended since we resolve it sooner.

Finally, we decided to calculate the projected baseline lengths required to detect deviation from circular symmetry at a 10\% level in the interferometric visibility. More specifically, using $|f-g|=0.1$, where $f$ and $g$ are the two curves in Figures~\ref{vishalpha} and~\ref{viskband} we determined the spatial frequency, $\theta$, at which there was 10\% difference between the two curves. The 10\% difference corresponded baselines of 150~m and 200~m for the H$\alpha$ and K-band regions, respectively.

\section{Discussion and Conclusion}
\label{sec4:conclusion}

The density structure and geometry of the spiral arms formed inside the disk of a Be binary system were studied in detail for the first time. Using our simulations from Paper 1, we examined the azimuthal dependence of the surface densities of the disk with radial distance. Gaussian functions were used to fit these surface density profiles in order to determine the azimuthal location and shape of the spiral arms. 

We investigated the effects of orbital phase, $p$, disk viscosity, $\alpha_\mathrm{SS}$, and the misalignment angle, $\theta$, on the winding parameter, $\gamma$, of the spiral arms. We find that $\gamma$ has only a small dependence on $p$, which is only present in misaligned systems. This is a good indication that the location of the secondary above and below the disk, in misaligned systems, can affect the structure of the spiral arms. We also find an inverse relationship between $\gamma$ and the $\alpha_\mathrm{SS}$ and $\theta$ parameters, i.e., an increase in either the viscosity of the disk or the misalignment angle of the orbit will result in more tightly wound spiral arms.

We investigated how the surface density of the arms is affected by phase, viscosity, and misalignment angle. We find that the trailing arm has little to no dependence on the orbital phase of the binary. Small differences in the outer parts of the leading arms were seen for misalignment angles of 30$\degr$ and 45$\degr$ which can be attributed to the elevation of the secondary above or below the equatorial plane. The viscosity of the disk is found to affect the density structure of the arms, specifically the surface density fall-off rate which is steeper for smaller $\alpha_\mathrm{SS}$ values. Finally, we find that the misalignment of the orbit affects mostly the outer parts of the spiral arms, where the surface density falls off more slowly in more misaligned systems, i.e. with increasing $\theta$. The radial distances at which these effects start to appear increases with viscosity. These distances are also consistently closer to the primary for trailing arms compared to the leading arms. 

Understanding the behaviour of these spiral arms is crucial to make sense of observations and to have a better grasp of the formation, evolution, and dissipation of Be star disks in both isolated and binary systems. For example, the presence of triple-peaked emission lines observed in the spectra of some Be stars may also be an indication of the presence of a binary companion. One such star, $\gamma$ Cas, is a well-studied early-type Be binary system that undergoes V/R variations and has been observed to display triply peaked H$\alpha$ profiles. \citet{nem12} suggest that changes within the density and extent of the disk may be responsible for these observed variations. \citet{esc15} also noted the presences of triple-peaked emission phases in the Be star $\zeta$ Tau. They speculate that these features occur within the outermost portions of the H$\alpha$ emitting region of the disk and may be due to the presence of a binary companion.

We also investigated whether these spiral density enhancements could be resolved with a typical interferometer and if these spiral arms would affect the axis ratios determined from interferometry. Specially, we illustrated how for our co-planar model a 10\% deviation from circular symmetry can be detected in the H$\alpha$ and K-band emitting regions at projected baselines of 150~m and 200~m, respectively, purely due to the presence of the disk density enhancements induced by a binary companion. Furthermore, we show that the axis ratios, as determined from typical Gaussian fits in these wavelength regimes can vary by as much as $\sim$20\% depending on interferometric baseline orientations. This could result in discrepancies when observations are obtained at different times when the location of the spiral density enhancements have changed position with respect to the observer. This investigation looked at one specific disk structure, a co-planar model with viscosity coefficient of 0.5. However, given that the departure from circular symmetry is related to the winding factor of the spiral arms, tighter spiral (smaller $\gamma$) would produce a more circular disk than a looser spiral (bigger $\gamma$). We also expect a large departure from circular symmetry for less viscous disks. Conversely we expect a smaller departure for more viscous systems as well as for non co-planar (misaligned) systems.   

In future work, we plan to investigate the evolution of these spiral features over time during disk building and dissipation phases to better understand the role of the binary companion in these processes. In particular, we plan to investigate the variability exhibited in emission lines with particular focus on changes in the V/R ratios during disk build-up and depletion phases to garner more insight into the formation of these systems. We also plan a follow-up study of the dynamics of individual or packets of particles in order to better understand how the motions within these disks affect the resulting features including the spiral arms.
\section*{Acknowledgements}

The authors thank the anonymous referee who's questions and comments helped improve the paper. This work was made possible by the facilities of the Shared Hierarchical Academic Research Computing Network (SHARCNET:www.sharcnet.ca), Compute/Calcul Canada, and Laboratory of Astroinformatics (IAG/USP, NAT/Unicsul, FAPESP grant No 2009/54006-4). CEJ acknowledges support from NSERC, the Natural Sciences and Engineering Research Council  program. CS recognizes support from the NSERC USRA program. C.\ T.\ acknowledges support from the National Science Foundation through grant AST-1614983. ACC and CEJ thanks UofT-FAPESP-UWO joint researcher program. ACC acknowledges support from CNPq (grant No. 311446/2019-1) and FAPESP (grant 2018/04055-8). The authors would also like to thank Keegan Marr for his help with some of plotting routines.

 \section*{Data availability} No new data were generated or analysed in support of this research.




\bibliographystyle{mnras}
\bibliography{main} %








\bsp	
\label{lastpage}
\end{document}